\documentclass[twocolumn,showpacs,preprintnumbers,amsmath,amssymb,pra]{revtex4}

\usepackage{amsmath}
\usepackage{amssymb}
\usepackage{graphicx}
\usepackage{dcolumn}
\usepackage{amsfonts}
\usepackage{amscd}

\newtheorem{thm}{Theorem}[section]
\newtheorem{lem}{Lemma}[section]
\newtheorem{prop}{Proposition}[section]
\newtheorem{cor}{Corollary}[section]

\newcommand{\lv}{\left \vert}
\newcommand{\rv}{\right \vert}
\newcommand{\la}{\left \langle}
\newcommand{\ra}{\right \rangle}
\newcommand{\ket}[1]{\lv #1 \ra}
\newcommand{\bra}[1]{\la #1 \rv}
\newcommand{\ketab}[1]{\lv #1 \ra_{AB}}
\newcommand{\keta}[1]{\lv #1 \ra_{A}}
\newcommand{\ketb}[1]{\lv #1 \ra_{B}}
\newcommand{\lV}{\left \Vert}
\newcommand{\rV}{\right \Vert}
\newcommand{\ip}[1]{\left\langle #1 \right\rangle}

\begin{document}
 
\title{Remote extraction and destruction of spread qubit information}

\author{Yoshiko Ogata}%
\affiliation{%
Graduate School of Mathematical Sciences,
The University of Tokyo, Tokyo, 153-8914 Japan\\
Department of Physics, The University of Tokyo, Tokyo 113-0033,
Japan
}%

\author{Mio Murao}
\affiliation{%
Department of Physics, The University of Tokyo, Tokyo
113-0033, Japan\\
PRESTO, JST, Kawaguchi, Saitama 332-0012, Japan\\
The Collaborative Institute for Nano Quantum Information
Electronics, The University of Tokyo, Tokyo 113-0033, Japan
}%

\begin{abstract}
Necessary and sufficient conditions for deterministic remote
extraction and destruction of qubit information encoded in bipartite
states using only local operations and classical communications
(LOCC) are presented. The conditions indicate that there is a way to
asymmetrically spread qubit information between two parties such
that it can be remotely extracted with unit probability at one of
the parties but not at the other as long as they are using LOCC.
Remote destruction can also be asymmetric between the two parties,
but the conditions are incompatible with those for remote
extraction.
\end{abstract}

\date{\today}


\maketitle

\section{Introduction}

Quantum information processing seeks to perform tasks which are
impossible or not efficient with the use of conventional classical
information processing, by using systems described by quantum
mechanics. We can consider two kinds of quantum information tasks
based on the types input states: the classical input tasks and the
quantum input tasks. Whereas input states of the classical input
tasks are quantum states but encode just classical information,
input states of the quantum input tasks encodes quantum information,
of which unit is described by a qubit $\alpha \ket{e_0} + \beta
\ket{e_1}$ where $\alpha$ and $\beta$ are unknown complex parameters
satisfying $\lv \alpha \rv^2+\lv \beta \rv^2=1$ and $\ket{e_0}$ and
$\ket{e_1}$ are a fixed basis of a qubit. For example, quantum
algorithms \cite{QuantumAlgorithms} are classical input tasks and
quantum error correcting codes \cite{QECC} and quantum universal
optimal cloning \cite{Qcloning} are quantum input tasks. To
investigate yet unveiled full quantum potential of quantum
information processing, it is necessary to understand properties of
quantum input tasks.

In many quantum input tasks, how quantum information is encoded in
the larger Hilbert space of composite systems determines the main
functionality of the tasks. For example, in quantum error correcting
codes, qubit information is encoded in a subspace of a larger
Hilbert space such that it can be still recovered after being
influenced by certain errors (or noises) which map input qubit
information into the whole Hilbert space.  The encoding process can
be described by a transformation of a computational basis $\{
\ket{i} \}$ where the original quantum information is given into a
set of orthogonal states in the larger Hilbert space $\{
\ket{\psi_i}\}$.  In this picture, the properties of encoding for a
task are captured by the choice of a set of states $\{
\ket{\psi_i}\}$, which represents {\it how} original quantum
information is spread across the Hilbert spaces of subsystems.

Entanglement, or a non-local quantum correlation, of an {\it
individual} state is an essential resource for performing quantum
input tasks such as quantum teleportation \cite{Teleportation},
namely, the existence of entanglement is necessary for performing
teleportation beyond the classical limit.  To analyze {\it
non-local} properties of spread quantum information described by the
set of states is a way to characterize how quantum information is
spread by encoding. Here, we use the word non-local to represent
properties which are not fully accessible by just using local
operations on the subspaces and classical communications (LOCC) but
global operations on the whole systems. For individual states, the
existence of this kind of non-locality is accompanied by the
existence of entanglement.

However, it is also known that such non-local properties of a set of
states can be essentially different from non-locality of individual
states. An important example is a set of nine product states which
cannot be locally discriminated by using LOCC presented in
\cite{Non-localityWithoutEntanglement} (the ``non-locality without
entanglement'' paper by Bennet et al.).  In this example, there is
no entanglement in the quantum states where classical information is
encoded, therefore, no entanglement resource required for encoding
classical information, but it is not possible to decode (i.e.,
identify encoded classical information) deterministically LOCC,
without using entanglement resources. Entanglement properties of
each encoded state does not fully capture the non-local property
appearing in the decoding process. As it had been also pointed out
in the context of local copy and local state discrimination in
\cite{LocalCopy-Owari}, impossibility of tasks involving LOCC
transformation of a set of states implies {\it non-locality beyond
individual entanglement}.

For characterizing non-locality of the spread of quantum
information, non-local resources required for decoding quantum
information should be considered as well as for encoding quantum
information, since required minimum resources for encoding process
and decoding process are not necessary the same in LOCC
transformations. In this paper, we focus on non-local resources
required for decoding processes and investigate properties of an
extreme case of spreading quantum information that does not consume
non-local resources for decoding. We study a simple but fundamental
case of spreading qubit information into two-party states. We
present necessary and sufficient conditions for spreading qubit
information into bipartite states such that qubit information can be
extracted by only using LOCC between the two parties.  We call this
task {\it remote extraction}.   In this task, since we have to
investigate simultaneous transformations of {\it two} states $\{
\ket{\psi_0}, \ket{\psi_1} \}$ under all possible LOCC, unlike the
case of a single known pure bipartite state where Lo-Popescu Theorem
\cite{Lo-Popescu} is applicable, the proof of necessity is involved
and it consists of seven steps. We also present the explicit form of
the LOCC for achieving the perfect remote extraction.

\begin{figure}
\includegraphics[width=75mm]{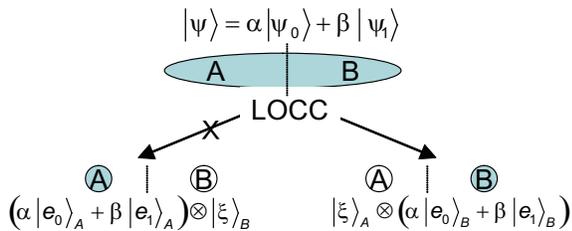}
\caption{Asymmetric remote extraction: We can encode single qubit
information into two parties where information can be extracted only
by using LOCC at Bob but not at Alice.
\label{fig:asymmetricLOCCextraction} }
\end{figure}

Interestingly, the obtained conditions indicate that qubit
information can be asymmetrically spread into a bipartite state.
(Fig.~\ref{fig:asymmetricLOCCextraction}). There is a way of
encoding to spread qubit information where it can be remotely
extracted with unit probability at one of the parties but not at the
other as long as they are using LOCC.  Thus, we can introduce a
non-local property of a set of states in an asymmetric manner for
two parties, whereas entanglement properties of an {\it individual}
bipartite pure state are always symmetric between the two parties
due to Schmidt decomposition. This fact is another indication of the
difference of non-locality of a set of states and an individual
state.

The asymmetric property of spreading quantum information can be used
for controlling the transmission of quantum information.  Namely, we
can distribute quantum information such that one of the party, say
Alice, can act as a controller of quantum information to support
other party to extract full quantum information, but she cannot
obtain full quantum information for herself, as long as they are
acting by LOCC.  Therefore we can consider a simple two qubit device
which globally encodes qubit information and also acts as a local
``switch'' for transmission of qubit information to one of the
qubits.

\begin{figure}
\includegraphics[width=75mm]{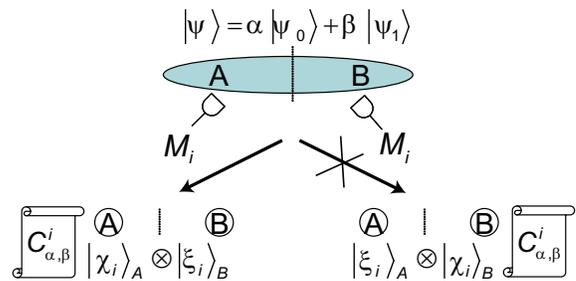}
\caption{Asymmetric remote destruction: Spread quantum information
can only be irreversibly destroyed by Alice's measurement on her
qubit, in a way such that the state after the measurement does not
contain quantum information ($\alpha$ and $\beta$).  Note that some
classical information represented by $C_{\alpha,\beta}^i$ can be
retrieved from the outcome of Alice's measurement $i$.
\label{fig:LOCCdestruction}}
\end{figure}

From the viewpoint of controlling transmission of quantum
information, it is also useful to spread quantum information into
two parties such that Alice's local operation can {\it irreversibly
destroy} quantum information such that the state of Bob's qubit
after Alice's operation is set to a {\it pure} state which does not
contain quantum information, i.e. the parameters $\alpha$ and
$\beta$. Note that this process is not a randomizing process to
transform Bob's qubit to be in a completely mixed state. In this
task, a part of classical information of the parameter $\alpha$ and
$\beta$ can be retrieved.  We call this task {\it remote
destruction}. (Fig.~\ref{fig:LOCCdestruction}) Such a way for
spreading quantum information can be used another kind of ``switch''
for controlling quantum information transmission. Using a similar
technique for proving conditions of remote extraction, we also
present necessary and sufficient conditions for this task. Spreading
quantum information for remote destruction can also be asymmetric
between two parties, but we show that the conditions for remote
destruction are incompatible for those of remote extraction.

This paper is organized as the following: In Section
\ref{statements}, the definitions and precise statements of remote
extraction and destruction are given. In Section
\ref{ext-sufficiency}, we show the proof of sufficiency for remote
extraction.  The preparations and outline of the proof of necessity
are given in Section \ref{ext-outline}.  The proof of necessity
consists of seven steps and they are presented in Section
\ref{ext-necessity}. The proof of conditions for remote destruction
is presented in Section \ref{des}, and the summary and discussions
are given in Section \ref{summary}.

\section{Statements of remote extraction and destruction}
\label{statements}

\subsection{Remote extraction}

We take two orthonormal vectors $\ketab{\psi_0},\ketab{\psi_1}$ in
two qubit Hilbert space ${\cal H}_{AB}={\mathbb C}^2\otimes {\mathbb
C}^2$, which we will call {\it basis states}, and encode qubit
information into a two-qubit state represented by
$\ket{\psi}=\alpha\ketab{\psi_0}+\beta\ketab{\psi_1}$. We consider
two qubits are spatially separated and one of the qubit is at
Alice's side and the other qubit is at Bob's side. The task of
remote extraction is to extract qubit information at Bob's side from
the two-qubit state $\ket{\psi}_{AB}$ by using finite rounds of
LOCC. That is, we look for a finite round LOCC procedure $\Lambda$
such that
\begin{widetext}
\begin{equation}
\Lambda \left(
\ketab{\alpha\psi_0+\beta\psi_1}\bra{\alpha\psi_0+\beta\psi_1}
\right) =\keta{\xi}\bra{\xi}\otimes\ketb{\alpha e_0+\beta e_1}
\bra{\alpha e_0+\beta e_1} \label{task}
\end{equation}
\end{widetext}
for arbitrary $\alpha,\beta\in{\mathbb C}$ satisfying $|\alpha
|^2+|\beta |^2=1$. Here, $\{\ketb{e_0}, \ketb{e_1}\}$ is a fixed
orthonormal basis in ${\cal H}_B$ and $\keta{\xi}$ is an arbitrary
vector in ${\cal H}_{A}$.  Throughout this paper, we use a notation
$\ket{\alpha\psi_0+\beta\psi_1} \equiv
\alpha\ket{\psi_0}+\beta\ket{\psi_1}$, and
$\bra{\alpha\psi_0+\beta\psi_1} \equiv \bar{\alpha}
\bra{\psi_0}+\bar{\beta}\bra{\psi_1}$ where $\bar{\alpha}$ and
$\bar{\beta}$ are complex conjugates of $\alpha$ and $\beta$,
respectively. We also denote the conjugation of a single qubit state
$\ket{\phi}= \alpha\ket{e_0}+\beta\ket{e_1}$ with respect to an
orthonormal basis $\{\ket{e_i}\}$ of the qubit by
$\ket{\bar{\phi}}=\bar{\alpha} \ket{e_0}+ \bar{\beta} \ket{e_1}$.

The finite round LOCC procedure $\Lambda$ is given by a sequence of
Alice's measurements
$\{M_{i_1,i_2,\cdots,i_k}^{j_1,\cdots,j_{k-1}}\otimes{\mathbb I} \}$
and Bob's $\{{\mathbb I}\otimes
N_{i_1,i_2,\cdots,i_k}^{j_1,\cdots,j_k}\}$, where $i_k$ is an index
for Alice's $k$-th round measurement and $j_k$ is an index for Bob's
$k$the round measurement ($k=1,\cdots,N$), satisfying the
normalization conditions
\begin{eqnarray}
\sum_{i_k}{M_{i_1,i_2,\cdots,i_k}^{j_1,\cdots,j_{k-1}}}^\dag
M_{i_1,i_2,\cdots,i_k}^{j_1,\cdots,j_{k-1}}=1,\\
\sum_{j_k}{N_{i_1,i_2,\cdots,i_k}^{j_1,\cdots,j_{k}}}^\dag
N_{i_1,i_2,\cdots,i_k}^{j_1,\cdots,j_{k}}=1
\end{eqnarray}
for each $k$. We use the notation $I_k=(i_1,\cdots,i_k),
J_k=(j_1,\cdots,j_k)$. It is easy to see that (\ref{task}) is
equivalent to
\begin{eqnarray}
({\mathbb I}\otimes
N_{I_N}^{J_N}) (M_{I_N}^{J_{N-1}}\otimes {\mathbb I}) \cdots
(M_{I_N}\otimes {\mathbb I})\ketab{\psi_i}=\keta{\xi}\ketb{e_i},
\label{eq}
\end{eqnarray}
for $i=0,1$ and all $(I_N,J_N)$. When there exists a LOCC procedure
satisfying the condition of Eq.(\ref{eq}), we say that qubit
information on the basis states $\ketab{\psi_0},\ketab{\psi_1}$ can
be {\it deterministically extracted} by LOCC at Bob's side.  We call
the pair of the state given by the right hand side of Eq.(\ref{eq})
$\{\keta{\xi}\ketb{e_0},\keta{\xi}\ketb{e_1}\}$ as the {\it
extracted form} and $\{\keta{\xi}v \ketb{e_0},\keta{\xi}v \ketb{
e_1}\}$ with some unitary operator $v$ is said to be {\it locally
equivalent to the extracted form to Bob}.  We use these notations in
our proof of remote extraction.

Now our problem is to find the condition of the basis states
$\{\ketab{\psi_0},\ketab{\psi_1}\}$ satisfying Eq.(\ref{eq}). In
this paper, we prove the following Theorem:
\begin{thm}\label{one}
Qubit information spread between Alice and Bob
$\ket{\psi}_{AB}=\alpha\ket{\psi_0}_{AB}+\beta\ket{\psi_1}_{AB}$ can
be deterministically extracted using only LOCC at Bob's side
$\ket{\phi}_B=\alpha\ket{e_0}_B+\beta\ket{e_1}_B$ if and only if the
Schmidt decompositions of the basis states
$\{\ket{\psi_0}_{AB},\ket{\psi_1}_{AB} \}$ are given by
\begin{eqnarray}
    \ket{\psi_0}_{AB}&=&\sqrt{\lambda_0} \ket{a_0}_A \ket{b_0}_B
    +\sqrt{\lambda_1} \ket{a_1}_A\ket{b_1}_B
    \label{cond1}
    \\
    \ket{\psi_1}_{AB}&=&\sqrt{\lambda_0} \ket{a_0'}_A \ket{b_1}_B
    +\sqrt{\lambda_1} \ket{a_1'}_A \ket{b_0}_B
    \label{cond2}
\end{eqnarray}
where $\lambda_i$ are the Schmidt coefficients satisfying $0 \le
\lambda_{1} \le \lambda_0 \le 1$ and $\lambda_0 + \lambda_1=1$, and
$\{\ket{a_i}\}$ and $\{\ket{a_i'}\}$ are the Schmidt basis of
Alice's qubit and $\{\ket{b_i}\}$ is the Schmidt basis of Bob's
qubit.  If the conditions of Eq.(\ref{cond1}) and (\ref{cond2}) are satisfied, then
 $\keta{a'_0},\keta{a'_1}$ are of the form
\begin{eqnarray}
\keta{a_0'} &=&  e^{-i\theta} \cos\Theta\keta{a_0} +\sin\Theta
\keta{a_1}, \\
\keta{a_1'}&=&e^{i\frac{\varphi}{2}}\left( -\sin\Theta \keta{a_0} +
e^{i \theta} \cos\Theta\ \keta{a_1}\right) \label{ov}
\end{eqnarray}
using three real parameters $\varphi$, $\theta$ and $\Theta$.

We define a family of orthonormal basis
$\{\keta{e_t^0},\keta{e_t^1}\}_{t\in{\mathbb R}_+}$ labeled by
positive real number $t$, by
\begin{eqnarray}
\keta{e^0_t}&=&\left( \mathcal{F}_{t,-\Theta} \keta{a_0}-i e^{i
\theta} \mathcal{F}_{-t,-\Theta} \keta{a_1}
\right)/\mathcal{N}_{t}\\
\keta{e^1_t}&=&\left( -i e^{- i \theta} \mathcal{F}_{-t,\Theta}
\keta{a_0}+ \mathcal{F}_{t,\Theta} \keta{a_1}
\right)/\mathcal{N}_{t}, \label{labelt}
\end{eqnarray}
where $\mathcal{F}_{t,\Theta}=1+t {\rm e}^{i \Theta}$ and $\mathcal{N}_{t}=\sqrt{2(t^2+1)}$.
Then if the pair of vectors $\{({\mathbb I}\otimes
N_{I_{k-1}}^{J_{k-1}})\cdots (M_{I_1}\otimes{\mathbb
I})\ketab{\psi_0},({\mathbb I}\otimes N_{I_{k-1}}^{J_{k-1}}) \cdots
(M_{I_1}\otimes{\mathbb I})\ketab{\psi_1}\}$ is not locally unitary
equivalent to the extracted form, all the measurements by Alice on
it are of the form
\begin{widetext}
\begin{eqnarray}
M^{J_{k-1}}_{I_k}=u_{I_k}^{J_{k-1}}
\left(\sqrt{\tau_{I_k,J_{k-1}}^0}
\ket{e^0_{t_{I_k,J_{k-1}}}}\bra{e^0_{t_{I_k,J_{k-1}}}}
+\sqrt{\tau_{I_k,J_{k-1}}^1}
\ket{e^1_{t_{I_k,J_{k-1}}}}\bra{e^1_{t_{I_k,J_{k-1}}}} \right)
\left( u^{J_{k-2}}_{I_{k-1}} \right)^\dagger, \label{measurement}
\end{eqnarray}
\end{widetext}
where $u_{I_k}^{J_{k-1}}$ is an unitary operator,
 $0\le \tau_{I_k,J_{k-1}}^0,\tau_{I_k,J_{k-1}}^1\le 1$
and $t_{I_k,J_{k-1}}\ge 0$.
On the other hand, all the measurements that
Bob carries out $\{N_{I_k}^{J_k}\}$ are
scalar multiplications of unitary operators.
\end{thm}

From the theorem, we see that the Schmidt coefficients of the basis
states have to be identical for remote extraction, therefore the
basis states should have, at least, same entanglement for remote
extraction. On the other hand, the Schmidt base of Alice's qubit of
the basis states are not necessary to be same.  Although the
orthogonality condition of the basis states $\bra{\psi_0} \psi_1
\rangle = \sqrt{\lambda_0 \lambda_1} ( \bra{a_1} a_0' \rangle +
\bra{a_0} a_1' \rangle ) =0$ fixes one of the parameters to be
$\varphi=0$, $\Theta=0$ or $\Theta=\pi$ for $\lambda_1 \neq 0$, we
can choose $\theta$ and one of $\varphi$ and $\Theta$ in
Eq.~(\ref{ov}) arbitrary.  This property allows asymmetry of remote
extraction; we can encode qubit information such that the conditions
for remote extraction at Bob are satisfied but the conditions for
remote extraction at Alice are not satisfied.

We can also obtain necessary and sufficient conditions of the basis
states for symmetric remote extraction, where the deterministic
remote extraction at either Alice or Bob is possible depending on
the choice of LOCC procedures from the Theorem 1.
\begin{cor}\label{both}
Extraction to either Alice or Bob is possible if and only if the
Schmidt decompositions of the basis states are given by
\begin{eqnarray}
\ketab{\psi_0}&=&\sqrt{\lambda_0} \keta{a_0}\ketb{b_0}
+\sqrt{\lambda_1} \ket{a_1}\ketb{b_1},\label{sym1}\\
\ketab{\psi_1}&=&-\sqrt{\lambda_0} \keta{a_1}\ketb{b_1}
+\sqrt{\lambda_1} \keta{a_0}\ketb{b_0}. \label{sym2}
\end{eqnarray}
\end{cor}

Thus, if a set of the basis states
$\{\ketab{\psi_0},\ketab{\psi_1}\}$ satisfies the conditions of
remote extraction Eqs.~(\ref{cond1}) and (\ref{cond2}), but does not
satisfy the conditions for the symmetric ones Eqs.~(\ref{sym1}) and
(\ref{sym2}), it gives an asymmetric way of spreading qubit
information, where deterministic remote extraction is only possible
at Bob, not at Alice.   For $\lambda_0=\lambda_1=1/\sqrt{2}$, any
choice of two orthogonal (maximally entangled) states can be
transformed into the forms of Eqs.~(\ref{cond1}) and (\ref{cond2}),
therefore there is no asymmetric remote extraction.  However, for
$\lambda_0 \neq \lambda_1$ where the Schmidt base are determined
uniquely, asymmetry of (perfect) remote extraction is guaranteed as
long as $\Theta \neq \pi/2$ in Eq.~(\ref{ov}).  The case of
$\lambda_0=1$ presents an interesting picture how qubit information is
spread between two parties in terms of symmetry and asymmetry; for
$\Theta=0$, qubit information is already extracted at Bob from the
beginning, and no qubit information can be extracted at Alice by
LOCC, for $\Theta=\pi/2$, qubit information is symmetrically shared
between Alice and Bob, and for $0<\Theta<\pi/2$, qubit information
is shared but asymmetrically.

We note that Bob's operation is restricted to scalar
multiplication of unitary operators for extracting qubit information
at Bob. Therefore, once one of the party performs an extraction
measurement of the form of Eq. (\ref{measurement}), qubit
information can be only extracted to the party who has not performed
the extraction measurement, even with the basis states allowing
symmetric remote extraction. The measurement condition also implies
that one-way LOCC, where Alice performs a projective measurement on
her qubit in the $\{\keta{e_t^0},\keta{e_t^1}\}$ basis and Bob
performs a conditional unitary operation depending on Alice's
measurement outcome is sufficient for remote extraction of qubit
information.

\subsection{Remote destruction}

The task of remote destruction is to {\it irreversibly} destroy
spread qubit information by acting one of the party (Alice) and to
prevent extracting quantum information at the other party (Bob). We
assume that Bob would not cooperate to destroy information, and also
we would like to prevent recovery of quantum information even if
classical information about Alice' measurement is known. We look for
Alice's measurement $\{M_i\}$ such that for arbitrary
$\alpha,\beta\in{\mathbb C}$ satisfying $|\alpha |^2+|\beta |^2=1$,
\begin{eqnarray}
(M_i\otimes 1)\ketab{ \alpha \psi_0 +\beta \psi_1}
=C^i_{\alpha,\beta}\keta{\chi_i}\ketb{\xi_i},\label{taskdis}
\end{eqnarray}
for each $i$.
Here, $C^i_{\alpha,\beta}$ is some scalar which depends on
$\alpha,\beta$, and $\keta{\chi_i},\ketb{\xi_i}$
are vectors that do not depend on $\alpha,\beta$.

In this paper, we show the following:
\begin{thm}\label{distract}
Deterministic remote destruction by Alice is possible if and only if
and only if the Schmidt decompositions of the basis states are given
by
\begin{eqnarray}
    \ket{\psi_0}_{AB}&=&\sqrt{\lambda_0} \ket{a_0}_A \ket{b_0}_B
    +\sqrt{\lambda_1} \ket{a_1}_A\ket{b_1}_B,\\
    \ket{\psi_1}_{AB}&=&\sqrt{\lambda_0} \ket{a_1}_A \ket{b_0'}_B
    +\sqrt{\lambda_1} \ket{a_0}_A \ket{b_1'}_B
\label{apple}
\end{eqnarray}
where $0\le\lambda_1 \le \lambda_0\le 1$, $\lambda_0+\lambda_1=1$
and $\{\ket{a_i}\}$ is the Schmidt base of Alice's qubit and
$\{\ket{b_i}\}$ and $\{\ket{b_i^\prime}\}$ are the Schmidt basis of
Bob's qubit. If the Schmidt rank of $\ketab{\psi_0}$
(resp.~$\ketab{\psi_1}$ ) is $2$, then the measurement operators for
deterministic remote destruction $\{M_i\}$ are of the form
\begin{eqnarray}
M_i=\ket{\chi_i}\bra{f_{k_i}}, \label{M}
\end{eqnarray}
where $\ket{\chi_i}$ is an arbitrary vector, $k_i=0,1$, and
$\{\ket{f_0},\ket{f_1}\}$ is an orthonormal basis diagonalizing a
matrix
\begin{widetext}
\begin{eqnarray*}
&(&\frac{1}{\sqrt{\lambda_0}}\ket{a_0}\bra{\bar{b}_0}
+\frac{1}{\sqrt{\lambda_1}}\ket{a_1}\bra{\bar{b}_1})
({\sqrt{\lambda_0'}}\ket{\bar{b'}_0}\bra{a_1}
+{\sqrt{\lambda_1'}}\ket{\bar{b'}_1}\bra{a_0} ) \\
(resp.
&(&\frac{1}{\sqrt{\lambda_0'}}\ket{a_1}\bra{\bar{b'}_0}
+\frac{1}{\sqrt{\lambda_1'}}\ket{a_0}\bra{\bar{b'}_1})
({\sqrt{\lambda_0}}\ket{\bar{b}_0}\bra{a_0}
+{\sqrt{\lambda_1}}\ket{\bar{b}_1}\bra{a_1} ) ).
\end{eqnarray*}
\end{widetext}

If the Schmidt rank of both of $\ketab{\psi_0}$ and $\ketab{\psi_1}$
are $1$, then the measurement operators $\{M_i\}$ are of the form
\[
M_i=\ket{\chi_i}\bra{\eta_i}.
\]
Here, the vector  $\ket{\eta_i}$ have to be $\ket{a_0}$ or
$\ket{a_0^\perp}$ if $\ketb{b_0}$ and $\ketb{b_0'}$ are not parallel to
each other, while it can be an arbitrary vector if
$\ketb{b_0}$ and $\ketb{b_0'}$ are parallel to each other.

\end{thm}

We see that the conditions given by Eq.~(\ref{apple}) is identical
for the conditions for deterministic remote extraction at {\it
Alice}, instead of Bob.  Therefore, the conditions for remote
destruction by Alice's measurement are incompatible for those of
remote extraction by Alice's measurement.  The conditions for
symmetric remote destruction are also given by Eqs.~(\ref{sym1}) and
(\ref{sym2}), therefore, in the symmetric case, Alice can determine
whether destructing qubit information or letting Bob to extract full
qubit information by the choice of her measurement, but Bob is also
in the same position.

\section{Proof of sufficiency for remote extraction}\label{ext-sufficiency}

We first observe that if the conditions given by Eqs.~(\ref{cond1})
and (\ref{cond2}) are satisfied, then the representation of the
base of Alice's qubit (Eq.~(\ref{ov})) is obtained.  The case of
$\lambda_0=0$ or $\lambda_1=1$ is trivial. Let us assume
$\lambda_0\lambda_1\neq0$.  Since $\ketab{\psi_0}$ and
$\ketab{\psi_1}$ are orthogonal, the two base of Alice's qubit
appearing in Eqs. (\ref{cond1}) and (\ref{cond2}) have to satisfy
\begin{eqnarray}
\langle a_0'\vert a_1\rangle +\langle a_1'\vert a_0\rangle=0.
\label{iti}
\end{eqnarray}
If we represent the basis state $\keta{a_0'}$ by $\keta{a_0'}=c_0
\keta{a_0}+c_1 \keta{a_1}$ and another basis state $\keta{a_1'}$ by
$\keta{a_1'}=e^{i\varphi} \left(-c^*_1 \keta{a_0}+c^*_0 \keta{a_1}
\right)$ with complex parameters $c_0$ and $c_1$ satisfying $\vert
c_0 \vert^2+\vert c_1 \vert^2=1$, and a real parameter $\varphi$,
the condition of Eq.(\ref{iti}) implies $c_1=c^*_1 e^{i\varphi}$. By
introducing two real parameters $\Theta$ and $\theta$, $c_0$ and
$c_1$ are represented by $c_0=\cos\Theta
e^{i(\frac{\varphi}{2}-\theta)}$ and $c_1=\sin\Theta
e^{i\frac{\varphi}{2}}$, respectively.  Thus, we obtain the
representation of the basis of Alice's qubit in Eq.~(\ref{ov}).

Now we choose another basis of Alice's qubit $\{ \ket{0}_A,
\ket{1}_A\}$ given by
\begin{eqnarray}
\keta{0}&=&\frac{1}{\sqrt 2}( \keta{a_0} -i{\rm e}^{i\theta}
\keta{a_1}
)\nonumber\\
\keta{1}&=&\frac{1}{\sqrt 2}(-i{\rm e}^{-i\theta} \keta{a_0}
+\keta{a_1}). \label{bs01}
\end{eqnarray}
We will check that qubit information can be extracted to Bob's qubit
by Alice's projective measurement described by $\{ \vert
0\rangle\langle 0\vert, \vert 1\rangle\langle 1\vert \}$ followed by
an appropriate unitary operation performed by Bob depending on the
measurement outcome of Alice.  If Alice obtains the measurement
result corresponding to $\keta{0}$, the basis states are transformed
to
\begin{eqnarray*}
\ketab{\psi_0}
\rightarrow \frac{1}{\sqrt 2}\keta{0} \ketb{ \sqrt{\lambda_0}b_0 +i{\rm
e}^{-i\theta}\sqrt{\lambda_1}b_1}
\end{eqnarray*}
\begin{eqnarray*}
\ketab{\psi_1} &\rightarrow
\frac{e^{i(\frac{\varphi}{2}+\Theta-\theta)}}{\sqrt 2} \keta{0}
\ketb{ \sqrt{\lambda_0} b_1 +ie^{i\theta}\sqrt{\lambda_1} b_0}.
\end{eqnarray*}
If Alice obtains the measurement result $\keta{1}$, the basis states
are transformed to
\begin{eqnarray*}
\ketab{\psi_0}
\rightarrow  \frac{1}{\sqrt 2}
\keta{1}
\ketb{
i{\rm e}^{i\theta}\sqrt{\lambda_0}b_0
+\sqrt{\lambda_1}b_1
},
\end{eqnarray*}
\begin{eqnarray*}
\ketab{\psi_1}
\rightarrow  \frac{1}{\sqrt 2}
e^{i(\frac{\varphi}{2}+\theta-\Theta)}
\keta{1}
\ketb{
i{\rm e}^{-i\theta}\sqrt{\lambda_0}
b_1
+\sqrt{\lambda_1}
b_0
}
\end{eqnarray*}
Note that the resulting pairs are locally equivalent to the
extracted form to Bob. Hence, by choosing a suitable unitary
operation transforming the basis of Bob's qubit back to $\{
\ket{e_0}, \ket{e_1}\}$, spread qubit information can be faithfully
extracted to Bob's side by only using LOCC.

\section{Preparations and outline for proof of necessity}
\label{ext-outline}

In our proof, we employ matrix representations of states. In this
section, we first describe the matrix representation, and then
introduce the key notion in our proof: {\it extraction measurements
(E-measurements)}. We also present the outline of our proof of
necessity for remote extraction consisting of seven steps.

\subsection{Matrix representation} Let
$\cal H$ be a $n$-dimensional Hilbert space, and let $\{\ket{f_i}
\}_{i=1}^{n}$ be an orthonormal basis of $\cal H$. We consider a
bi-partite system ${\cal H}\otimes{\cal H}$. Let
$\ketab{\Omega}=\sum_{i=1}^{n}\frac{1}{\sqrt n}\keta{f_i}\ketb{f_i}$
be a maximal entangled state in ${\cal H}\otimes{\cal H}$. The
conjugation of a state $\ket{\xi}=\sum_i\alpha_i \ket{f_i}\in {\cal
H}$ with respect to $\{\ket{f_i}\}$ is represented by $
\ket{\bar\xi} =\sum_i\bar\alpha_i \ket{f_i}\in {\cal H}$. The
conjugation of an operator $X\in B({\cal H})$ with respect to a basis
$\{\ket{f_i} \}_{i=1}^{n}$ is denoted by $\bar X$, i.e.,
\begin{eqnarray*}
X=\sum_{ij}\beta_{ij}\vert f_i\rangle\langle f_j\vert \to \bar
X=\sum_{ij}\bar\beta_{ij}\vert f_i\rangle\langle f_j\vert.
\end{eqnarray*}
One can easily check that the useful relations $\bar X \ket{\bar\xi}
=\ket{\overline{ X \xi}}$, $(\vert\eta\rangle\langle\xi\vert \otimes
1)\ketab{\Omega} =\frac{1}{\sqrt
n}\vert\eta\rangle_A\vert\bar\xi\rangle_B$ and
$\langle\bar\xi\vert\bar\eta\rangle =\langle\eta\vert\xi\rangle
=\overline{\langle\xi\vert\eta\rangle}$.  These relations are
extensively used in our proof.

By straight forward calculation, we can check the following properties:
\begin{prop}\label{form}
\quad\\
\begin{enumerate}
\item For all $\ketab{\psi}\in{{\cal H}\otimes{\cal H}}$,
there exists unique $X\in B({\cal H})$ such that
$\ket{\psi}=(X\otimes 1)\ketab{\Omega}$.
\item
$\langle\Omega,(X\otimes1)\Omega \rangle
=\frac{1}{n}Tr X.$
\item
$(X\otimes 1)\ketab{\Omega}=(1\otimes\bar X^\dagger)\ketab{\Omega}$
\end{enumerate}
\end{prop}

\subsection{Extraction measurements}

An E-measurement performed by Alice on a pair of orthonormal states
$\{\ketab{\psi_0},\ketab{\psi_1}\}$ is a measurement described by a
set of measurement operators $\{M_i\otimes {\mathbb I}\}$ satisfying
$\sum_i M_i^\dag M_i={\mathbb I}$, which preserve orthogonality of
the states $\langle {\psi_0} \vert (M_i^\dagger \otimes {\mathbb I})
(M_i\otimes {\mathbb I}) \vert {\psi_1}\rangle=0$ for all $i$ and
also equi-probability, namely, equal probability for measuring each
basis state $\Vert(M_i\otimes {\mathbb I})\ketab{\psi_0}\Vert
=\Vert(M_i\otimes {\mathbb I})\ketab{\psi_1}\Vert$, while there
exists $i$ such that $M_i^\dag M_i\neq{\mathbb R}_+{\mathbb I}$. An
E-measurement performed by Bob is defined in the same manner. An
E-measurement is not always possible and the existence of the
E-measurement restricts the form of
$\{\ketab{\psi_0},\ketab{\psi_1}\}$. Note that the final pair of
extraction $\{\keta{\xi}\ketb{e_0},\keta{\xi}\ketb{e_1}\}$ is
measurable by E-measurement (E-measurable) of Alice given by
$\{\ket{\xi}\bra{\xi}, \ket{\xi^\perp}\bra{\xi^\perp}\}$. On the
other hand, we call another type of measurement such that $M_i^\dag
M_i\in{\mathbb R}_+ {\mathbb I}$ for all $i$, a {\it C-measurement}.
Note that if extraction to Bob is possible, Alice should be able to
perform the E-measurement on the last pair, otherwise extraction to
Bob at the next round is not possible.

Now, we introduce a set of orthonormal base of Alice's qubit,
$S_A(\ketab{\psi_0},\ketab{\psi_1})$. We define
$S_A(\ketab{\psi_0},\ketab{\psi_1})$ by a set of all orthogonal
basis $\{\keta{0},\keta{1}\}$ such that the decompositions
\begin{eqnarray}
\ketab{\psi_0}&=&\keta{0}\ketb{\xi}+\keta{1}\ketb{\eta},\nonumber\\
\ketab{\psi_1}&=&\keta{0}\ketb{\xi^\perp}+\keta{1}\ketb{\eta^\perp},
\label{bunk}
\end{eqnarray}
satisfy
\begin{eqnarray}
\Vert\xi \Vert=\Vert { \xi^{\perp}}\Vert,
\Vert {
\eta} \Vert=\Vert { \eta^{\perp}}\Vert,
\bra{\xi}\xi^{\perp}\rangle=\bra{\eta} \eta^{\perp}\rangle=0.
\label{deco}
\end{eqnarray}
Of course, $S_A(\ketab{\psi_0},\ketab{\psi_1})$ can be an empty set,
depending on $\{ \ketab{\psi_0},\ketab{\psi_1} \}$. We call an
element in $S_A(\ketab{\psi_0},\ketab{\psi_1})$, an orthonormal
basis on which Alice can perform an E-measurement. In fact, we will
see that if Alice can operate an E-measurement $\{M_i\}$ on
$\{\ketab{\psi_0},\ketab{\psi_1}\}$, then each $M_i$ have to be of
the form
\begin{eqnarray}
M_i=\sqrt{\tau_i^0}u_i\keta{0} \bra{0}
+\sqrt{\tau_i^1}u_i\keta{1}\bra{1}, \label{mform}
\end{eqnarray}
where $u_i$ is a single qubit unitary, $\{\keta{0},\keta{1}\}\in
S_A(\ketab{\psi_0},\ketab{\psi_1})$, and $0\le \tau_i^0,\tau_i^1\le
1$.  We also define $S_B(\ketab{\psi_0},\ketab{\psi_1})$ in the same
manner. Then it is obvious that for arbitrary single qubit unitary
operators $u,v$ and a complex number $c\neq 0$, we have
\begin{eqnarray}
S_A\left(\left(cu\otimes v\right)\ketab{\psi_0},\left(cu\otimes
v\right)\ketab{\psi_1}\right) = uS_A\left(\ketab{\psi_0},
\ketab{\psi_1}\right). \nonumber
\end{eqnarray}

\subsection{Outline of proof}

We prove the necessary conditions for remote extraction in the
following seven steps:
\begin{description}
\item{\it Step 1:}
We prove that the orthogonality and equi-probability conditions
should be satisfied for all rounds of LOCC. From this, we show that
the local operations in the LOCC procedure have to be E-measurements
or C-measurements.
\item{\it Step 2:}
We show that if Alice can perform an E-measurement $\{M_i\}$ on a
pair
 $\{\ketab{\psi_0},\ketab{\psi_1}\}$,
then $S_A(\ketab{\psi_0},\ketab{\psi_1})$ is non-empty. Furthermore,
we see that each $M_i$ has to be of the form given by
Eq.~(\ref{mform}).
\item{\it Step 3:}
We derive the explicit form of the set
$S_A(\ketab{\psi_0},\ketab{\psi_1})$ when it is not empty. We see
that it is parameterized by a positive scalar $t\ge0$.
\item{\it Step 4:}
We derive the necessity conditions for both of Alice and Bob to be
able to perform an E-measurement on a pair of states
$\{\ketab{\psi_0},\ketab{\psi_1}\}$.
\item{\it Step 5:}
Using the result of {\it Step 4}, we prove that the following
situation is impossible: Alice performs some E-measurement $\{M_i\}$
on $\{\ketab{\psi_0},\ketab{\psi_1}\}$, and for all the results of
her measurement $\{(M_i \otimes{\mathbb I})\ketab{\psi_0}, (M_i
\otimes{\mathbb I})\ketab{\psi_1}\}_i$, Bob can sequently perform an
E-measurement.
\item{\it Step 6:}
We show that if deterministic remote extraction is possible,
 $S_A(\ketab{\psi_0},\ketab{\psi_1})$ is not empty.
\item{\it Step 7:}
We show that Eqs.~(\ref{bunk}) and (\ref{deco}) imply that the
Schmidt forms of $\ketab{\psi_0},\ketab{\psi_1}$ to be given by
Eqs.~(\ref{cond1}) and (\ref{cond2}).
\end{description}

\section{Proof for necessity of remote extraction}\label{ext-necessity}

\subsection{{\it Step 1}: Orthogonality and equi-probability }

We show that deterministic remote extraction requires that the two
vectors (unnormalized  basis states) have to be orthogonal to each
other and have the same norm at every step in LOCC. Let us consider
a LOCC described by a sequence of conditional local measurements of
Alice $\{M_{i_1,i_2,\cdots,i_k}^{j_1,\cdots,j_{k-1}}\otimes{\mathbb
I} \}$ and Bob $\{{\mathbb I}\otimes
N_{i_1,i_2,\cdots,i_k}^{j_1,\cdots,j_k}\}$, for $k=1,\cdots,N$. Each
set of measurement operators $\{M_{i_1,\cdots,i_{k+1}}^{j_1,\cdots
j_{k}}\}$, $\{N_{i_1,\cdots,i_{k}}^{j_1,\cdots j_{k}}\}$ satisfies
\begin{eqnarray}
    \sum_{i_{k+1}}
    \left( M_{i_1,\cdots,i_{k+1}}^{j_1,\cdots j_{k}}
    \right)^\dagger
    M_{i_1,\cdots,i_{k+1}}^{j_1,\cdots j_{k}}
    ={\mathbb I},\nonumber\\
    \sum_{j_k}
    \left(
    N_{i_1,\cdots,i_{k}}^{j_1,\cdots j_{k}}
    \right)^\dagger
    N_{i_1,\cdots,i_{k}}^{j_1,\cdots j_{k}}
    ={\mathbb I}.
\label{sum}
\end{eqnarray}
We use a notation $I_k=(i_1,\cdots,i_k), J_k=(j_1,\cdots,j_k)$, as
introduced in Section~\ref{statements}, and denote the vectors at
each step by
\begin{eqnarray*}
\ketab{\psi_{0}^{I_k,J_{m}}}=
\left(
M_{I_k}^{J_{k-1}}\cdots M_{I_1}
\otimes
N_{I_m}^{J_m}\cdots N_{I_1}
\right)
\ketab{\psi_0}.\\
\ketab{\psi_{1}^{I_k,J_{m}}}=
\left(
M_{I_k}^{J_{k-1}}\cdots M_{I_1}
\otimes
N_{I_m}^{J_m}\cdots N_{I_1}\right)
\ketab{\psi_1},
\end{eqnarray*}
where $m=k-1$ or $m=k$. As seen in Section~\ref{statements}, at the
last turn ($k=N$), the two vectors are orthogonal $\
    \langle
\psi_{0}^{I_N,J_{N}} \vert \psi_{1}^{I_N,J_{N}} \rangle
    =0$ and they have the same length $\Vert \vert \psi_{0}^{I_N,J_{N}} \rangle \Vert
    =
    \Vert \vert \psi_{1}^{I_N,J_{N}} \rangle \Vert$.

By summing up with respect to $j_N$, using the relation (\ref{sum}),
we have $\langle \psi_{0}^{I_N,J_{N-1}} \vert \psi_{1}^{I_N,J_{N-1}}
\rangle
    =0$ and $\Vert \vert \psi_{0}^{I_N,J_{N-1}} \rangle \Vert
    =
    \Vert \vert \psi_{1}^{I_N,J_{N-1}} \rangle \Vert$.
Repeating this summation procedure, we obtain $\langle
\psi_{0}^{I_k,J_{m}} \vert \psi_{1}^{I_k,J_{m}} \rangle
    =0$ and $\Vert \vert \psi_{0}^{I_k,J_{m}} \rangle \Vert
    =
    \Vert \vert \psi_{1}^{I_k,J_{m}} \rangle \Vert$
for all $k=1,\cdots, N$ and $m=k-1,k$, i.e., the orthogonality and
equi-probability conditions should be satisfied for all rounds in
LOCC. Therefore, the local operations in the LOCC procedure have to
be E-measurements or C-measurements. As the C-measurements can not
extract information on its own, we need the E-measurements.

\subsection{{\it Step 2}: E-measurement by Alice} We derive
the necessity and sufficient conditions for Alice to be able to
carry out the E-measurement.

\begin{lem}\label{on}
If Alice can carry out an E-measurement on a pair of orthonormal
states $\{\ketab{\psi_0},\ketab{\psi_1}\}$ in ${\cal H}_{AB}$, then,
$S_A(\ketab{\psi_0},\ketab{\psi_1})$ is not empty. Furthermore, the
E-measurement have to be of the form (\ref{mform}).
\end{lem}

{\it Proof:} Let $\{M_i\otimes {\mathbb I}\}_i$ be an E-measurement
by Alice on $\{\ketab{\psi_0},\ketab{\psi_1}\}$. As it is the
E-measurement, there exists $i$ such that $M_i^\dagger M_i\neq
{\mathbb R_+ \mathbb I}$. As $M_i^\dagger M_i$ is positive, it can
be diagonalized in a suitable basis $\{\keta{0},\keta{1}\}$. We will
show that $\{\keta{0},\keta{1}\}\in
S_A(\ketab{\psi_0},\ketab{\psi_1})$. In the basis
$\{\keta{0},\keta{1}\}$, we have
\begin{eqnarray*}
M_i^\dagger M_i=
\begin{pmatrix}
\tau_0&0\\
0&\tau_1
\end{pmatrix},
\end{eqnarray*}
where $0\le \tau_0<\tau_1 \le 1$.  We define two matrices $X_0$ and
$X_1$ for the matrix representation of the basis states
$\ketab{\psi_0}=(X_0 \otimes 1) \ket{\Omega}_{AB}$ and
$\ketab{\psi_1}=(X_1 \otimes 1) \ket{\Omega}_{AB}$. Let us represent
$X_0,X_1$ in this basis $ \{ \keta{0},\keta{1} \}$ as
\begin{eqnarray*}
X_0=
\begin{pmatrix}
a&b\\
c&d
\end{pmatrix}\quad
X_1=
\begin{pmatrix}
x&y\\
z&w
\end{pmatrix}.
\end{eqnarray*}
As $\ketab{\psi_0},\ketab{\psi_1}$ are orthogonal unit vectors
satisfying  $\ip{\psi_0 \vert \psi_1}=0$ and
$\lV{\ket{\psi_0}}\rV=\lV{\ket{\psi_1}}\rV$, we have
\begin{eqnarray*}
(\vert a\vert^2+\vert b\vert^2)+(\vert c\vert^2+\vert d\vert^2)
&=&(\vert x\vert^2+\vert y\vert^2)+(\vert z\vert^2+\vert w\vert^2)\\
(a\bar x+b\bar y)+(c\bar z+d\bar w)&=&0
\end{eqnarray*}
The condition of the Proposition is represented in this basis as
$
\tau_0(\vert a\vert^2+\vert b\vert^2)
+\tau_1(\vert c\vert^2+\vert d\vert^2)
=\tau_0(\vert x\vert^2+\vert y\vert^2)
+\tau_1(\vert z\vert^2+\vert w\vert^2)$ and
$\tau_0(a\bar x+b\bar y)+\tau_1(c\bar z+d\bar w)=0$.
Hence we have
\begin{eqnarray}
\vert a\vert^2+\vert b\vert^2
&=&\vert x\vert^2+\vert y\vert^2\\
\vert c\vert^2+\vert d\vert^2
&=&\vert z\vert^2+\vert w\vert^2 \\
a\bar x+b\bar y &=&c\bar z+d\bar w=0. \label{alph}
\end{eqnarray}
These conditions are rewritten in term of $\{\keta{0},\keta{1}\}$ as
follows: We have
\begin{eqnarray*}
X_0 \ket{\Omega}&=&(\vert 0\rangle\langle 0\vert+ \vert
1\rangle\langle\vert 1\vert)X_0 \ket{\Omega} =\vert 0\rangle\vert
\xi\rangle
+\vert 1\rangle\vert \eta\rangle, \\
X_1 \ket{\Omega} &=&(\vert 0\rangle\langle 0\vert+ \vert
1\rangle\langle 1\vert)X_1 \ket{\Omega} =\vert 0\rangle\vert
\xi^\perp\rangle +\vert 1\rangle\vert \eta^\perp\rangle.
\end{eqnarray*}
where
\begin{eqnarray*}
\xi&=&\frac{1}{\sqrt 2}\overline{X_0^\dagger\vert 0\rangle},\quad
\xi^\perp=\frac{1}{\sqrt 2}\overline{X_1^\dagger\vert 0\rangle},\\
\eta&=&\frac{1}{\sqrt 2}\overline{X_0^\dagger\vert 1\rangle},\quad
\eta^\perp=\frac{1}{\sqrt 2}\overline{X_1^\dagger\vert 1\rangle}.
\end{eqnarray*}
It is easy to check that Eq.~(\ref{alph}) is equivalent to
\begin{eqnarray}
\bra{\xi}\xi^{\perp}\rangle=\bra{\eta} \eta^{\perp}\rangle=0
,~\Vert\xi \Vert=\Vert { \xi^{\perp}}\Vert, \Vert { \eta}
\Vert=\Vert { \eta^{\perp}}\Vert \label{vr}
\end{eqnarray}
and we conclude the basis $\{\keta{0},\keta{1}\}$ is in
$S_A(\ketab{\psi_0},\ketab{\psi_1})$. As $\{\keta{0},\keta{1}\}$ was
defined as a basis that diagonalizes $M_i^\dagger M_i$, $M_i$ has to
be of the form given by Eq.~(\ref{mform}), i.e., the E-measurement
have to be of the form of Eq.~(\ref{mform}). $\square$

\subsection{{\it Step 3}: The set $S_A(\ketab{\psi_0},\ketab{\psi_1})$}

We derive the explicit form of vectors in the set
$S_A(\ketab{\psi_0},\ketab{\psi_1})$.

\begin{lem}
Suppose that $\{\ketab{\psi_0}\ketab{\psi_1}\}$ is not local
unitary equivalent to the extracted form, and $S_A(\ketab{\psi_0},\ketab{\psi_1})$ is not empty.
Let us fix one element $\{\keta{0},\keta{1}\}$
in $S_A(\ketab{\psi_0},\ketab{\psi_1})$.
Then
\begin{eqnarray*}
S_A(\ketab{\psi_0},\ketab{\psi_1})
=\{\keta{e^0_t},\keta{e^1_t}\}_{t\ge 0},
\end{eqnarray*}
where $\{\keta{e^0_t},\keta{e^1_t}\}$ is an orthonormal basis of
${\cal H}_A$, labeled by a positive real number $t$:
\begin{eqnarray}
\keta{e^0_t}&=&\frac{1}{\sqrt{t^2+1}}(\keta{0}+te^{i\zeta}\keta{1}),\\
\keta{e^1_t}&=&\frac{1}{\sqrt{t^2+1}}(-te^{-i\zeta}\keta{0}+\keta{1}).
\label{rt}
\end{eqnarray}
Here, the phase factor $e^{i\zeta}$ is determined as follows:
If the
Schmidt rank of $\ketab{\psi_0}$ is $2$, then we have
$\langle\eta\vert\xi^\perp\rangle\neq 0$,
$\vert\frac{\langle\xi\vert\eta^\perp\rangle}{\langle\eta\vert\xi^\perp\rangle}\vert=1$,
and we define a phase factor $e^{2i\zeta}$ by
$e^{2i\zeta}=-\frac{\langle\xi\vert\eta^\perp\rangle}{\langle\eta\vert\xi^\perp\rangle}$.
If the Schmidt rank of $\ketab{\psi_0}$ is $1$ and
$\{\ketab{\psi_0},\ketab{\psi_1}\}$ is not local unitary
equivalent to the extracted form, then we have
$\langle\xi\vert\eta\rangle\neq\langle\xi^\perp\vert\eta^\perp\rangle$,
and we define the phase
$e^{2i\zeta}=-\frac{\langle\xi\vert\eta\rangle
-\langle\xi^\perp\vert\eta^\perp\rangle}
{\langle\eta\vert\xi\rangle
-\langle\eta^\perp\vert\xi^\perp\rangle}$.
\end{lem}

{\it Proof}\\
From the equivalence of Eqs.~(\ref{alph}) and (\ref{vr}), the matrix
forms of the basis states, $X_0$ and $X_1$, are represented by
\begin{eqnarray}
X_0=\begin{pmatrix}
a&b\\
c&d
\end{pmatrix},\quad
X_1=
\begin{pmatrix}
-\bar b&\bar a\\
-{\rm e}^{2i\zeta}\bar d&{\rm e}^{2i\zeta}\bar c
\end{pmatrix}
\label{mr}
\end{eqnarray}
in the $\{\keta{0},\keta{1}\}$ basis. A general orthonormal basis
can be written as $\ket{{e_\kappa^0}}= ( \keta{0} +\kappa \keta{1})
/ \sqrt{1+\vert \kappa \vert^2}$ and $\ket{{e_\kappa^1}}= (
-\bar\kappa \keta{0} + \keta{1} ) / \sqrt{1+\vert \kappa \vert^2}$
in terms of $\{ \keta{0}, \keta{1} \}$ using a parameter $\kappa \in
{\mathbb C}$. If $\{\keta{e^0_\kappa},\keta{e^1_\kappa}\}\in
S_A(\ketab{\psi_0},\ketab{\psi_1})$, then the two basis states have
to satisfy the condition $\bra{{e_\kappa^i}} X_0 X_1^\dag
\ket{{e_\kappa^i}}=0$ and $\Vert X_0^\dag \ket{{e_\kappa^i}} \Vert
=\Vert X_1^\dag \ket{{e_\kappa^i}} \Vert$, for $i=0,1$. By
Eq.~(\ref{mr}), this condition is equivalent to
\begin{eqnarray*}
\vert \bar a+\bar c\kappa \vert^2+\vert \bar b+\bar d\kappa \vert^2
=\vert b+{\rm e}^{-2i\zeta}d\kappa\vert^2
+\vert a +{\rm e}^{-2i\zeta}c\kappa\vert^2\\
(a+ c\bar \kappa)(-b-{\rm e}^{-2i\zeta}d\kappa)
+(b+ d\bar \kappa)(a+{\rm e}^{-2i\zeta}c\kappa)=0
\end{eqnarray*}
These conditions are also equivalent to the following conditions
\begin{eqnarray*}
(\kappa-{\rm e}^{2i\zeta}\bar\kappa)(a \bar c+b \bar d
-\bar b d{\rm e}^{-2i\zeta}
-\bar a c{\rm e}^{-2i\zeta})=0\\
(\kappa-\bar\kappa{\rm e}^{2i\zeta})
(ad-bc)=0
\end{eqnarray*}
If
\begin{equation}
ad\neq bc \label{net1}
\end{equation}
or
\begin{equation}
\bar{a} c+\bar b d-b\bar d{\rm e}^{2i\zeta} -a\bar c{\rm
e}^{2i\zeta}\neq 0, \label{net2},
\end{equation}
we have $\kappa=\bar\kappa{\rm e}^{2i\zeta}$, hence we obtain
$\kappa= t{\rm e}^{i\zeta}$ where $ t\in{\mathbb R}$. However, it is
easy to see $ \{\keta{e^0_t},\keta{e^1_t}\} =\{\keta{e^0_{-\frac
1t}},\keta{e^1_{-\frac 1t}}\}$ for $t>0$. Therefore, if (\ref{net1})
or (\ref{net2}) holds, we can parameterize
$S_A(\ketab{\psi_0},\ketab{\psi_1})$ with a positive scalar
$t\in{\mathbb R}_+$. It is also easy to see that if the Schmidt rank
of $\ketab{\psi_0}$ is $2$, (\ref{net1}) holds. Furthermore, in this
case, $e^{2i\zeta}$ is given by
$e^{2i\zeta}=-\frac{\langle\xi\vert\eta^\perp\rangle}{\langle\eta\vert\xi^\perp\rangle}$.
On the other hand, if the Schmidt rank of $\ketab{\psi_0}$ is $1$
and it is not local unitary equivalent to the extracted form, then
Eq.~(\ref{net2}) holds. In this case, the phase is given by
$e^{2i\zeta}=-\frac{\langle\xi\vert\eta\rangle
-\langle\xi^\perp\vert\eta^\perp\rangle} {\langle\eta\vert\xi\rangle
-\langle\eta^\perp\vert\xi^\perp\rangle}$. Hence, in both cases,
$S_A(\ketab{\psi_0},\ketab{\psi_1})$ is $t$-parameterized. $\square$

By a direct calculation, complex conjugation of the vectors
$\ketb{\xi_t},\ketb{\xi_t^\perp},\ketb{\eta_t},\ketb{\eta_t^\perp}$
in
\begin{eqnarray*}
\ketab{\psi_0}&=&\vert e_t^1\rangle_A\vert\xi_t^1\rangle_B
+\vert e_t^2\rangle_A\vert\xi_t^2\rangle_B,\\
\ketab{\psi_1}&=&\vert e_t^1\rangle_A\vert\xi_t^{1\perp}\rangle_B
+\vert e_t^2\rangle_A\vert\xi_t^{2\perp}\rangle_B
\end{eqnarray*}
are given by
\begin{eqnarray}
\ketb{\overline{\xi_t}}&=&\frac{1}{\sqrt{2(1+t^2})}
\begin{pmatrix}
\bar a+{\rm e}^{i\zeta}t\bar c \\
\bar b+{\rm e}^{i\zeta}t\bar d
\end{pmatrix},\\
\ketb{\overline{\xi_t^{\perp}}}&=&\frac{1}{\sqrt{2(1+t^2)}}
\begin{pmatrix}
-b-{\rm e}^{-i\zeta}td\\
a+{\rm e}^{-i\zeta}t c
\end{pmatrix}\\
\ketb{\overline{\eta_t}}&=&
\frac{1}{\sqrt{2(1+t^2)}}
\begin{pmatrix}
-\bar at{\rm e}^{-i\zeta}+\bar c\\
-\bar bt{\rm e}^{-i\zeta}+\bar d
\end{pmatrix},\\
\ketb{\overline{\eta_t^{\perp}}}
&=&\frac{1}{\sqrt{2(1+t^2)}}
\begin{pmatrix}
tb{\rm e}^{-i\zeta}-{\rm e}^{-2i\zeta}d\\
-a t{\rm e}^{-i\zeta} +{\rm e}^{-2i\zeta} c
\end{pmatrix},
\label{xi}
\end{eqnarray}
in the $\{\ketb{0},\ketb{1}\}$ basis .

\subsection{{\it Step 4}\; :\; E-operation from both sides}
Suppose that both of Alice and Bob can perform an E-measurement on a
pair of basis states $\{\ketab{\psi_0},\ketab{\psi_1}\}$. This
assumption excludes the possibility that
$\{\ketab{\psi_0},\ketab{\psi_1}\}$ is local unitary equivalent to
the extracted form at Alice or Bob from the beginning. Since Alice
can perform an E-measurement, $S_A(\ketab{\psi_0},\ketab{\psi_1})$
is non-empty and its elements are $t$-parameterized
$S_A(\ketab{\psi_0},\ketab{\psi_1})
=\{\keta{e^0_t},\keta{e^1_t}\}_{t\ge 0}$ as we have shown in {\it
Step 3}. The vectors $\{\ketab{\psi_0},\ketab{\psi_1}\}$ can be
decomposed with respect to the elements of
$S_A(\ketab{\psi_0},\ketab{\psi_1})$
\begin{eqnarray}
\ketab{\psi_0}&=& \keta{e_t^0}\ketb{\xi_t}+\keta{e_t^1}\ketb{\eta_t},\\
\ketab{\psi_1}&=& \keta{e_t^0}\ketb{\xi_t^\perp}
+\keta{e_t^1}\ketb{\eta_t^\perp},
\label{tB}
\end{eqnarray}
so that $\bra{\xi_t}\xi_t^{\perp}\rangle=\bra{\eta_t}
\eta_t^{\perp}\rangle=0$ and $\Vert { \xi_t} \Vert=\Vert {
\xi_t^{\perp}}\Vert, \Vert { \eta_t} \Vert=\Vert {
\eta_t^{\perp}}\Vert$. Furthermore, every E-measurement by Alice on
$\{\ketab{\psi_0},\ketab{\psi_1}\}$ is of the form
\begin{eqnarray}
M_i=\sqrt{\tau_i^0}u_i\keta{e_{t_i}^0}\bra{e_{t_i}^0}
+\sqrt{\tau_i^1}u_i\keta{e_{t_i}^1}\bra{e_{t_i}^1},
\label{mf}
\end{eqnarray}
where $u_i$ is a unitary operator, $0\le \tau_i^0,\tau_i^1\le 1$ and
$t_i\in{\mathbb R}_+$. In {\it Step 4}, we show that under the
assumption that both Alice and Bob can perform an E-measurement on
$\{\ketab{\psi_0},\ketab{\psi_1}\}$, the vectors
$\ketb{\xi_t},\ketb{\xi_t^\perp},\ketb{\eta_t}$ and
$\ketb{\eta_t^\perp}$ satisfy $\langle\xi_t\vert\eta_t\rangle_B+
\langle\xi_t^{\perp}\vert\eta_t^{\perp}\rangle_B=0$ and $\Vert
{\xi_t} \Vert=\Vert {\eta_t} \Vert$ for all $t\ge0$.

To prove this, note that if Bob can perform an E-measurement, there
exists a basis set $\{ \ketb{e_0'},\ketb{e_1'} \}\in
S_B(\ketab{\psi_0},\ketab{\psi_1})$ satisfying
\begin{eqnarray}
\langle \overline{e_i'}\vert X_0^\dagger X_0  \vert \overline{e_i'}\rangle&=&\langle
\overline{e_i'}\vert X_1^\dagger X_1\vert \overline{e_i'}\rangle,
\nonumber\\
\langle
\overline{e_i'}\vert X_0^\dagger X_1\vert \overline{e_i'}\rangle&=&0,
\label{ao}
\end{eqnarray}
for $i=0,1$, from {\it Step 2}.  Note that $X_0,X_1$ can be
represented as
\begin{eqnarray*}
X_0&=&\sqrt{2}(\ket{e^0_t}\bra{\bar{\xi_t}}
+\ket{e^1_t}\bra{\bar{\eta_t}}\\
X_1&=&\sqrt{2}(\ket{e^0_t}\bra{\bar{\xi_t^\perp}}
+\ket{e^1_t}\bra{\bar{\eta_t^\perp}}) ,
\end{eqnarray*}
because of Eq.~(\ref{tB}). Let us represent the $t$-parameterized
vectors in the $\{ \keta{\overline{e_0'}},\keta{\overline{e_1'}} \}$
basis:
\begin{eqnarray*}
\ket{\bar{\xi_t}}&=&\alpha_t^0\ket{\overline{e_0'}}+\beta_t^0\ket{\overline{e_1'}},\\
\ket{\bar{\eta_t}}&=&\alpha_t^1\ket{\overline{e_0'}}+\beta_t^1\ket{\overline{e_1'}},\\
\ket{\bar{\xi_t^{\perp }}}&=&{\rm e}^{i\varphi^0_t}
(-\bar{({\beta}_t^0)}\ket{\overline{e_0'}}+\bar{({\alpha}_t^0)}\ket{\overline{e_1'}}),\\
\ket{\bar{\eta_t^{\perp }}}&=&{\rm e}^{i\varphi^1_t}
(-\bar{({\beta}_t^1)}\ket{\overline{e_0'}}+\bar{({\alpha}_t^1)}\ket{\overline{e_1'}}).
\end{eqnarray*}
Then $X_0^\dagger X_0,X_1^\dagger X_1,X_0^\dagger X_1$ are
represented in the $\{ \keta{e_0'},\keta{e_1'} \}$ basis as
\begin{eqnarray*}
X_0^\dagger X_0&=&\sum_i
\begin{pmatrix}
\vert\alpha^i_t\vert^2&*\\
*&\vert\beta^i_t\vert^2
\end{pmatrix},\\
X_1^\dagger X_1&=&\sum_i
\begin{pmatrix}
\vert\beta^i_t\vert^2&*\\
*&\vert\alpha^i_t\vert^2
\end{pmatrix}\\
X_0^\dagger X_1 &=&\sum_i
{\rm e}^{i\varphi^i_t}
\begin{pmatrix}
-\bar\beta_i^t\bar\alpha^i_t&*\\
*&\bar\alpha_i^t\bar\beta^i_t
\end{pmatrix}
\end{eqnarray*}
where $*$ represents irrelevant elements for our evaluation. Hence
Eq.~(\ref{ao}) implies
\begin{eqnarray*}
\sum_i\vert\alpha^i_t\vert^2=\sum_i\vert\beta^i_t\vert^2\\
\sum\alpha^i_t\beta^i_t{\rm e}^{-i\varphi^i_t}=0
\end{eqnarray*}
It is easy to derive the relations
$\vert\beta^1_t\vert=\vert\alpha^0_t\vert$ and
$\vert\beta^0_t\vert=\vert\alpha^1_t\vert$ which imply $\Vert \xi_t
\Vert=\Vert \eta_t\Vert$ for all $t\ge 0$.

Representing $X_0,X_1$ in the $\{\keta{0},\keta{1}\}$ basis, from
the representation in  Eq.~(\ref{xi}), we see
\begin{widetext}
\begin{eqnarray}
&\Vert\xi_t\Vert=\Vert\eta_t\Vert\nonumber\\
&\Leftrightarrow
t^2\left(\vert c\vert^2+\vert d\vert^2-\vert a\vert^2-\vert b\vert^2
\right)
+2t\left(
{\rm e}^{-i\zeta}c\bar a
+{\rm e}^{i\zeta}\bar ca
+{\rm e}^{-i\zeta}d\bar b
+{\rm e}^{i\zeta}\bar d b
\right)
-(\vert c\vert^2+\vert d\vert^2-\vert a\vert^2-\vert b\vert^2)=0
\nonumber\\
&\Leftrightarrow
(t^2-1)(\Vert {\eta_0} \Vert^2- \Vert {\xi_0}
\Vert^2)+2t{\rm e}^{-i\zeta}(\langle\xi_0\vert\eta_0\rangle_+
\langle\xi^{\perp}_0\vert\eta_0^{\perp}\rangle)=0,
\label{te}
\end{eqnarray}
\end{widetext}
for all $t$. This implies $\langle\xi_0\vert\eta_0\rangle+
\langle\xi_0^{\perp}\vert\eta_0^{\perp}\rangle=0$.  But as we have a
freedom about the choice of the fixed basis $\{\keta{0},\keta{1}\}$
(we could take $\{\keta{0}=\keta{e_t^0}, \keta{1}=\keta{e_t^1}\}$),
we obtain $\langle\xi_t\vert\eta_t\rangle+
\langle\xi_t^{\perp}\vert\eta_t^{\perp}\rangle=0$ for all $t\ge 0$.
In the matrix representation in Eq. (\ref{xi}), we have
\begin{eqnarray}
\langle\xi_t\vert\eta_t\rangle+
\langle\xi_t^{\perp}
\vert\eta_t^{\perp}\rangle=0\nonumber\\
\Leftrightarrow
(-t^2+1)(a\bar c{\rm e}^{2i\zeta}+b\bar d{\rm e}^{2i\zeta}
+d\bar b+\bar ac) \nonumber \\
+2{\rm e}^{i\zeta}t\left(
\vert c\vert^2+\vert d\vert^2-\vert a\vert^2-\vert b\vert^2
\right)
=0.
\label{to}
\end{eqnarray}

\subsection{{\it Step 5}\; : \; Impossibility of sequent
E-measurement} Let us consider the following situation: Alice
performs some E-measurement $\{M_i\}$ on
$\{\ketab{\psi_0},\ketab{\psi_1}\}$, and for all the results of the
measurements $\{(M_i\otimes{\mathbb I} )\ketab{\psi_0}, (
M_i\otimes{\mathbb I})\ketab{\psi_1}\}_i$, Bob can sequently perform
another E-measurement. Can this situation occur? In this {\it Step
5}, we show this is not possible. By symmetry, the situation that
interchanging Alice's and Bob's roles is also impossible.

If this situation occurs, the pair
$\{\ketab{\psi_0},\ketab{\psi_1}\}$ is not local unitary equivalent
to the extracted form at Bob. Therefore,
$S_A(\ketab{\psi_0},\ketab{\psi_1})$ should be $t$-parameterized and
$\{\ketab{\psi_0},\ketab{\psi_1}\}$ is decomposed as in
Eq.~(\ref{tB}). Each $M_i$ is of the form of Eq. (\ref{mf}). As it
is an E-measurement, there exists $i$ such that
$\tau_i^0\neq\tau_i^1$. After Alice's E-measurement, the two basis
states are transformed as
$
\ketab{\psi_0^i}=(M_i\otimes{\mathbb I} )\ketab{\psi_0}
=\sqrt{\tau_{i}^0}
u_i\keta{e_{t_i}^0}\ketb{\xi_{t_i}}
    +\sqrt{\tau_i^1}u_i\keta{ e_{t_i}^1}\ketb{\eta_{t_i}}$ and
$\ketab{\psi_1^i}=(M_i\otimes{\mathbb I})\ketab{\psi_1}
=\sqrt{\tau_i^0}u_i\keta{ e_{t_i}^0}
\ketb{\xi_{t_i}^{\perp}}
    +\sqrt{\tau_i^1}u_i\keta{e_{t_i}^1}\ketb{\eta_{t_i}^{\perp}}$.
Note that Alice still can perform an E-measurement on this pair,
(with $\{\keta{e_{t_i}^0}\bra{e_{t_i}^0}$,
$\keta{e_{t_i}^1}\bra{e_{t_i}^1}\}$, for example.) Now assume that
Bob can perform an E-measurement on
$\ketab{\psi_0^i},\ketab{\psi_1^i}$, for all $i$. Then, we have
$\tau_i^0\tau_i^1\neq 0$, and the pair of basis states
$\ketab{\psi_0^i},\ketab{\psi_1^i}$ have to satisfy
\begin{eqnarray}
\sqrt{\tau_i^0}
\Vert{\xi_{t_i}}\Vert =\sqrt{\tau^1_i}\Vert {\eta_{t_i}}
\Vert,\label{af}\\
\left\langle  \xi_{t_i} \vert
\eta_{t_i} \right\rangle +\left\langle  \xi_{t_i}^{\perp} \vert
\eta_{t_i}^{\perp}\right\rangle =0
\label{ef}
\end{eqnarray}
by {\it Step 4}. As $\{M_i\}$ is an E-measurement, there exists $i$
such that $\tau_i^0\neq \tau_i^1$. In the following, we see that if
there exists $i$ such that $\tau_i^0\neq\tau_i^1$, Eqs.~(\ref{af})
and (\ref{ef}) have at most one solution $t_i=t'$ in ${\mathbb
R}^+$.

Note that Eq.~(\ref{ef}) is equivalent to Eq.~(\ref{to}). If $a\bar
c{\rm e}^{2i\zeta}+b\bar d{\rm e}^{2i\zeta} +d\bar b+\bar ac=0$, and
$\vert c\vert^2+\vert d\vert^2-\vert a\vert^2-\vert b\vert^2=0$ are
satisfied, we have $\Vert\xi_t\Vert=\Vert\eta_t\Vert$ for all $t$
from Eq.~(\ref{te}). From Eq.~(\ref{af}), this implies
$\tau_i^0=\tau^1_i$ for all $i$, which contradicts our assumption.
Therefore, we have $a\bar c{\rm e}^{2i\zeta}+b\bar d{\rm
e}^{2i\zeta} +d\bar b+\bar ac\neq0$ or $\vert c\vert^2+\vert
d\vert^2-\vert a\vert^2-\vert b\vert^2\neq 0$.
If $a\bar c{\rm e}^{2i\zeta}+b\bar d{\rm e}^{2i\zeta} +d\bar b+\bar
ac=0$ and $\vert c\vert^2+\vert d\vert^2-\vert a\vert^2-\vert
b\vert^2\neq 0$ are satisfied, then Eq.~(\ref{ef}) has the only
solution $t=0$.
If $a\bar c{\rm e}^{2i\zeta}+b\bar d{\rm e}^{2i\zeta} +d\bar b+\bar
ac\neq0$,  we have
$
t^2-1
-2{\rm e}^{i\zeta}t\left(
\vert c\vert^2+\vert d\vert^2-\vert a\vert^2-\vert b\vert^2
\right)(a\bar c{\rm e}^{2i\zeta}+b\bar d{\rm e}^{2i\zeta}
+d\bar b+\bar ac)^{-1}
=0$.
This equation has one negative solution and one positive solution.
Hence in any case, Eqs.~(\ref{af}) and (\ref{ef}) have at most one
solution in${\mathbb R}_+$.

If Eqs.~(\ref{af}) and (\ref{ef}) have no solution, we can conclude
that the situation is impossible. Let us consider the case that
there exists a unique solution
 $t=s\ge 0$.
As each $i$ has to satisfy Eqs.~(\ref{af}) and (\ref{ef}), we have
$t_i=s$ for all $i$. Then by Eq.~(\ref{af}), we obtain
${\tau_i^0}/{\tau_i^1}={\Vert
\keta{\eta_{s}}\Vert^2}/{\Vert\keta{\xi_{s}}\Vert^2 } \equiv r$ for
all $i$, i.e., the ratio of $\tau_i^0$ and $\tau_i^1$ is independent
of $i$. Furthermore, as there exists $i$ such that
$\tau_i^0\neq\tau_i^1$, the ratio $r$ is not $1$. Therefore, we have
\begin{eqnarray*}
\sum_iM_i^\dagger M_i&=&
\tau^0_i\vert e_{t_i}^0\rangle\langle e_{t_i}^0\vert
+\tau^1_i\vert e_{t_i}^1\rangle\langle e_{t_i}^1\vert \\
&=&\frac{1}{s^2+1}
\sum_i(\tau_i^1)
\begin{pmatrix}
r+s^2&(r
-1)s{\rm e}^{-i\zeta}\\
(r
-1)s{\rm e}^{i\zeta}&
rs^2+1
\end{pmatrix}.
\end{eqnarray*}
As for $r \neq 1$, this is not equal to $\mathbb I$. This
contradicts the normalization condition of the measurement operator
$
\sum_iM_i^\dagger M_i={\mathbb I}.
$
Therefore, the situation we have considered cannot occur.

\subsection{{\it Step 6}\; :\; Necessary conditions for
remote extraction} Suppose that remote extraction to Bob's qubit is
possible somehow.
If $\{\ketab{\psi_0},\ketab{\psi_1}\}$ is not local unitary
equivalent to the extracted form, it should be possible to carry out
the first E-measurement, for either Alice or Bob. If Alice can carry
it out, $S_A(\ketab{\psi_0},\ketab{\psi_1})$ is not empty. On the
other hand, if Alice can not perform the first E-measurement, Bob
have to be able to do it. However, it is impossible from the
following reason: Recall that the final pair Eq.~(\ref{eq}) is a
pair of the basis states that Alice can carry out an E-measurement.
As we consider only finite rounds of LOCC, this means at some point
of LOCC, Bob performs an E-measurement and after any result of Bob's
E-measurement, Alice should be able to perform an E-measurement
i.e., the situation considered in {\it Step 5} should occur.
However, we have shown that it is impossible in {\it Step 5}.
Therefore, the extraction to Bob's qubit is only available for the
case that Alice can perform the first E-measurement, i.e., the case
that $S_A(\ketab{\psi_0},\ketab{\psi_1})$ is not empty. Furthermore,
as we have seen above, Bob can carry out only C-measurements in the
extraction procedure.

\subsection{Schmidt picture}
In this {\it Step 7}, we represent the conditions of
Eq.~(\ref{bunk}) for $S_A(\ketab{\psi_0},\ketab{\psi_1})$ in the
Schmidt form given by Eqs. (\ref{cond1}) and (\ref{cond2}). Suppose
that Eq.~(\ref{bunk}) is satisfied and let
\begin{eqnarray*}
\ket{\psi_0}_{AB}=\sqrt{\lambda_0} \ket{a_0}_A \ket{b_0}_B
    +\sqrt{\lambda_1} \ket{a_1}_A\ket{b_1}_B\\
\ket{\psi_1}_{AB}=\sqrt{\lambda_0'} \ket{a_0'}_A \ket{b_0'}_B
    +\sqrt{\lambda_1'} \ket{a_1'}_A \ket{b_1'}_B
\end{eqnarray*}
be the Schmidt decompositions of
$\ketab{\psi_0},\ketab{\psi_1}$, respectively.
Then we have
\begin{eqnarray*}
X_0/\sqrt{2} &=&\sqrt{\lambda_0}\vert a_0
\rangle\langle \bar{b_0}\vert +\sqrt{\lambda_1}
\vert a_1\rangle\langle
\bar{b_1}\vert =\vert 0\rangle\langle \bar{\xi}\vert +\vert 1\rangle\langle
\bar{\eta}\vert\\
X_1/\sqrt{2} &=&\sqrt{\lambda_0'}\vert a_0'\rangle\langle
\bar{b_0'}\vert +\sqrt{\lambda_1'}\vert a_1'\rangle\langle  \bar{b_1'}\vert =
\vert 0\rangle\langle \bar{\xi^{\perp}}\vert +\vert 1\rangle\langle
\bar{\eta^{\perp }}\vert.
\end{eqnarray*}
By the relations $\bra{\xi}
\xi^{\perp}\rangle=\bra{\eta}\eta^{\perp}\rangle=0$ and
$\Vert\ket{\xi}\Vert=\Vert\ket{\xi^{\perp}}\Vert$,
$\Vert\ket{\eta}\Vert=\Vert\ket{\eta^{\perp}}\Vert$, the vectors
$\ketb{\xi},\ketb{\xi^\perp},\ketb{\eta},\ketb{\eta^\perp}$ are
represented in the basis parallel to $\ketb{\xi}$ as
\begin{eqnarray*}
\ketb{\xi}&=&
\begin{pmatrix}
s\\0
\end{pmatrix}
,\quad
\ketb{\xi^\perp}=
\begin{pmatrix}
0\\s
\end{pmatrix},\\
\ketb{\eta}&=&
\begin{pmatrix}
\alpha' \\
\beta'
\end{pmatrix},\quad
\ketb{\eta^\perp}= e^{i\Lambda}
\begin{pmatrix}
-\bar\beta'\\
\bar\alpha'
\end{pmatrix},
\end{eqnarray*}
for $s\in{\mathbb R}$ and $\alpha', \beta'\in{\mathbb C}$.
Therefore, we have
\begin{eqnarray*}
\vert \xi\rangle\langle \xi\vert
+\vert \eta\rangle\langle \eta\vert
=
\begin{pmatrix}
s^2+\vert\alpha'\vert^2&\alpha'\bar\beta'\\
\beta'\bar\alpha'&\vert\beta'\vert^2
\end{pmatrix}
\\
=
\lambda_0\ketb{b_0}\bra{b_0}+\lambda_1\ketb{b_1}\bra{b_1}
\\
\vert \xi^\perp\rangle\langle \xi^\perp\vert
+\vert \eta^\perp\rangle\langle \eta^{\perp}\vert
=
\begin{pmatrix}
\vert\beta'\vert^2&-\alpha'\bar\beta'\\
-\beta\bar\alpha'&s^2+\vert\alpha'\vert^2
\end{pmatrix}
\\
=
\lambda_0'\ketb{b_0'}\bra{b_0'}+\lambda_1'\ketb{b_1'}\bra{b_1'}
\end{eqnarray*}
From these relations, we derive that
$\lambda_0=\lambda_0',\lambda_1=\lambda_1'$, $\ket{b_0'}$ is
parallel to $\ket{b_1}$, and $\ket{b_1'}$ is parallel to
$\ket{b_0}$. By modifying the phase of $\ket{a'_0}$ and $\ket{a_1'}$
appropriately, we can take $\ket{b_0'}=\ket{b_1}$ and
$\ket{b_1'}=\ket{b_0}$.

\subsection{The form of LOCC}
Finally, we present the explicit form of LOCC. By Section
\ref{ext-sufficiency}, the basis $\{\keta{0},\keta{1}\}$ represented
by
$
\keta{0}=(
\keta{a_0}-i e^{i \theta}\keta{a_1})
/\sqrt 2$ and
$\keta{1}=( -i e^{-i \theta}
\keta{a_0}+ \keta{a_1}
)/\sqrt{2}$
is an element of $S_A(\ketab{\psi_0},\ketab{\psi_1})$. Therefore, if
$\{\ketab{\psi_0},\ketab{\psi_1}\}$ is not local unitary equivalent
to the extracted form, all the elements in
$S_A(\ketab{\psi_0},\ketab{\psi_1})$ are $t$-parameterized by {\it
Step 3}, with respect to $\{\ket{0}_A,\ket{1}_A\}$, which is equal
to the form of Eq.~(\ref{labelt}) in the Theorem.

By {\it Step 5}, Bob can carry out only a C-measurement for remote
extraction at Bob, therefore, all $N_{I_k}^{J_k}$ are scalar
multiplications of unitary operators. On the other hand, the form of
Alice's measurement on the pair of the states
$\{\ketab{\psi_0},\ketab{\psi_1}\}$ have to be
$M_i=\sqrt{\tau_i^0}u_i\ket{e_{t_i}^0}\bra{e_{t_i}^0}
+\sqrt{\tau_i^1}u_i\ket{e_{t_i}^1}\bra{e_{t_i}^1}$, with a unitary
operation $u_i$, $0\le\tau_i^0,\tau_i^1\le 1$ and
$\{\ket{e_{t_i}^0},\ket{e_{t_i}^1}\}\in
S_A(\ketab{\psi_0},\ketab{\psi_1})$. For this $M_i$ and a nonzero
scalar multiplication of a unitary operator $N_i$, we can see
$S_A(M_i\otimes N_i\ketab{\psi_0}, M_i\otimes
N_i\ketab{\psi_1})=u_iS_A(\ketab{\psi_0},\ketab{\psi_1})$ if
$\tau_i^0\tau_i^1\neq 0$, from the argument in {\it Step 2}. Note
that if $\tau_i^0\tau_i^1=0$, the pair $\{M_i\otimes
N_i\ketab{\psi_0}, M_i\otimes N_i\ketab{\psi_1}\}$ is locally
unitary equivalent to the extracted form at Bob. Thus, we obtain
the form of Alice's measurement given in the Theorem \ref{one},
inductively.

\section{Proof for Remote Destruction}\label{des}
Let $\ketab{\psi_0},\ketab{\psi_1}\in{\cal H}_{AB}$, $\la \psi_0
\vert \psi_1\ra=0$. We consider the basis states in the matrix
representation: $\ketab{\psi_0}=(X_0\otimes 1)
\ket{\Omega}_{AB},\ketab{\psi_1}=(X_1\otimes 1)\ket{\Omega}_{AB}$.
First we show the following lemma.
\begin{lem}\label{odis}
Suppose that there exists a matrix $M\neq 0$ and vectors
$\ket{\chi}$, $\ket{\xi}$ satisfying the following property: for all
$\alpha, \beta\in{\mathbb C}$, there exists
$C_{\alpha,\beta}\in{\mathbb C}$ such that
\[
(M\otimes 1)\left(\alpha\ketab{\psi_0}+\beta\ketab{\psi_1}\right)
=C_{\alpha,\beta}\keta{\chi}\otimes\ketb{\xi}.
\]
Then if the Schmidt rank of $\ketab{\psi_0}$
(resp. $\ketab{\psi_1}$)
is $2$,
\[
M=\ket{\chi}\bra{\eta},
\]
where $\ket{\eta}$ is an eigenvector of a matrix
$({X_0}^\dagger)^{-1}{X_1}^\dagger$ (resp. $({X_1}^\dagger)^{-1}{X_0}^\dagger$).

If the Schmidt rank of both of $\ketab{\psi_0}$ and $\ketab{\psi_1}$
are $1$ and the basis states are represented by
\[
\ketab{\psi_0}=\keta{f}\ketb{\xi},
\ketab{\psi_1}=\keta{f'}\ketb{\xi'},
\]
then one of the followings occurs:
\begin{enumerate}
\item $M=\ket{\chi}\bra{f^\perp}$
\item $M=\ket{\chi}\bra{{f'}^{\perp}}$
\item $\ketb{\xi}$ and $\ketb{\xi'}$ are parallel to each other
and $M=\ket{\chi}\bra{\eta}$ for an arbitrary vector $\ket{\eta}$.
\end{enumerate}
\end{lem}

{\it Proof}\\
If the Schmidt rank of $\ket{\psi_0}$ is $2$, and $ (M\otimes
1)\ketab{\psi_0} =C_{1,0}\keta{\chi}\otimes\ketb{\xi}$, $M$ has to
be of rank $1$. We represent it as $M=\ket{\chi}\bra{\eta}$ by introducing a vector
$\ket{\eta}$. Then we have
\begin{eqnarray*}
(M\otimes 1)\ketab{\psi_0} &=&\left( \ket{\chi}\bra{{X_0}^\dagger\eta}
\otimes 1\right)\Omega \\
=\frac{1}{\sqrt{2}}
\keta{\chi}\ketb{(\overline{{X_0}^\dagger\eta})}
&=&C_{1,0}\keta{\chi}\otimes\ketb{\xi},\\
(M\otimes 1)\ketab{\psi_1} &=&\left( \ket{\chi}\bra{{X_1}^\dagger\eta}
\otimes 1\right)\Omega \\
=\frac{1}{\sqrt{2}}
\keta{\chi}\ketb{(\overline{{X_1}^\dagger\eta})}
&=&C_{0,1}\keta{\chi}\ketb{\xi},
\end{eqnarray*}
Hence we obtain
\begin{eqnarray*}
\ket{{X_0}^\dagger\eta}=\sqrt{2}\overline{C_{1,0}}  \ket{\bar{\xi}},
\ket{{X_1}^\dagger\eta}=\sqrt{2}\overline{C_{0,1}}  \ket{\bar{\xi}}.
\end{eqnarray*}
As the Schmidt rank of $\ketab{\psi_0}$ is $2$,
 the rank of ${X_0}$ is $2$, i.e., ${X_0}$ is invertible.
Therefore, $C_{1,0}\neq 0$, and
\begin{eqnarray*}
\sqrt{2}\ket{\bar{\xi}}
=\frac{1}{\overline{C_{1,0}}}\ket{{X_0}^\dagger\eta}.
\end{eqnarray*}
Then we obtain
\begin{eqnarray*}
\ket{{X_1}^\dagger\eta}=\frac{\overline{C_{0,1}}}{\overline{C_{1,0}}}
\ket{{X_0}^\dagger\eta}.
\end{eqnarray*}
As ${X_0}$ is invertible, we have
\begin{eqnarray*}
\ket{({X_0}^\dagger)^{-1}{X_1}^\dagger\eta}
=\frac{\overline{C_{0,1}}}{\overline{C_{1,0}}}
\ket{\eta}.
\end{eqnarray*}
Hence $\ket{\eta}$ is an eigenvector of
$({X_0}^\dagger)^{-1}{X_1}^\dagger$.  The result is also unchanged for the
case that Schmidt rank of $\ket{\psi_1}$ is $1$.

On the other hand, suppose that the Schmidt rank of both basis
states $\ketab{\psi_0}$ and $\ketab{\psi_1}$ are $1$. We write
$\ketab{\psi_0}=\keta{f}\ketb{\xi},\ketab{\psi_1}=\keta{f'}\ketb{\xi'}$.
By the assumption, $\keta{Mf}\ketb{\xi}$ and $\keta{Mf'}\ketb{\xi'}$
have to be parallel. From this fact, the statement holds.
$\square$\\\\

Next we present the following Lemma for the cases where the Schmidt
the Schmidt rank of $\ketab{\psi_0}$ (resp.$\ketab{\psi_1}$ ) is
$2$.
\begin{lem} \label{matrixform1}
If the Schmidt rank of $\ketab{\psi_0}$ (resp.$\ketab{\psi_1}$ ) is
$2$, deterministic remote destruction is possible if and only if the
matrix $ (\frac{1}{\sqrt{\lambda_0}}\ket{a_0}\bra{\bar{b}_0}
+\frac{1}{\sqrt{\lambda_1}}\ket{a_1}\bra{\bar{b}_1})
({\sqrt{\lambda_0'}}\ket{\bar{b'}_0}\bra{a_0'}
+{\sqrt{\lambda_1'}}\ket{\bar{b'}_1}\bra{a_1'} ) $ (resp.$
(\frac{1}{\sqrt{\lambda_0'}}\ket{a_0'}\bra{\bar{b'}_0}
+\frac{1}{\sqrt{\lambda_1'}}\ket{a_1'}\bra{\bar{b'}_1})
({\sqrt{\lambda_0}}\ket{\bar{b}_0}\bra{a_0}
+{\sqrt{\lambda_1}}\ket{\bar{b}_1}\bra{a_1} ) $) is diagonalized by
some orthonormal basis $\{\ket{f_0},\ket{f_1}\}$ with eigenvalues
$z_0, z_1 \in{\mathbb C}$.

Furthermore, the measurement
operators for deterministic remote destruction $\{M_i\}$ are of the
form
\begin{eqnarray}
M_i=\ket{\chi_i}\bra{f_{k_i}}, \label{M2}
\end{eqnarray}
where $\ket{\chi_i}$ is an arbitrary vector and $k_i=0,1$.
\end{lem}

{\it Proof:} Note that
\begin{widetext}
\begin{eqnarray*}
({X_0}^\dagger)^{-1}{X_1}^\dagger
=2\left(\frac{1}{\sqrt{\lambda_0}}\ket{a_0} \bra{\bar{b}_0}
+\frac{1}{\sqrt{\lambda_1}}\ket{a_1} \bra{\bar{b}_1}\right) \left(
{\sqrt{\lambda_0'}}\ket{\bar{b}'_0} \bra{a_0'}
+{\sqrt{\lambda_1'}}\ket{\bar{b}'_1} \bra{a_1'} \right).
\end{eqnarray*}
\end{widetext}
If $({X_0}^\dagger)^{-1}{X_1}^\dagger$ is diagonalized by an
orthonormal basis $\{ \ket{f_k} \}$ with corresponding eigenvalues
$\{ z_k \}$, then we have
${X_1}^\dagger\ket{f_k}=z_k{X_0}^\dagger\ket{f_k}, $ for $k=0,1$. We
set $ \ket{\xi^k}\equiv\ket{\overline{{X_0}^\dagger f_k}}$. By
defining $M_0=\ket{f_0}\bra{f_0},M_1=\ket{f_1}\bra{f_1}$, we obtain
\begin{widetext}
\begin{eqnarray*}
(M_k\otimes 1)\ketab{\psi_0}=(M_k{X_0}\otimes 1) \ket{\Omega}
=(\ket{f_k}\bra{{X_0}^\dagger f_k}\otimes 1) \ket{\Omega}
=\frac{1}{\sqrt{2}}\keta{f_k}\ketb{\overline{{X_0}^\dagger f_k}}
=\frac{1}{\sqrt{2}}\keta{f_k}\ketb{\xi^k},\\
(M_k\otimes 1)\ketab{\psi_1}=(M_k{X_1}\otimes 1)\ket{\Omega}
=(\ket{f_k}\bra{{X_1}^\dagger f_k}\otimes 1) \ket{\Omega}
=\frac{1}{\sqrt{2}}\keta{f_k}\ketb{\overline{{X_1}^\dagger f_k}}
=\frac{2}{\sqrt{2}}\bar{z_k}\keta{f_k}\ketb{\xi^k}.
\end{eqnarray*}
\end{widetext}
Hence we have
$
(M_k\otimes 1)\left(\alpha\ketab{\psi_0}+\beta\ketab{\psi_1}\right)
=\left((\alpha+2\beta\bar{z_k})/ \sqrt{2} \right)
\keta{f_k}\ketb{\xi^k},
$
and remote destruction is possible.

Conversely, suppose that a measurement represented by $\{M_i\}_i$ of
deterministic remote destruction is possible. Then, from Lemma
\ref{odis}, we have $M_i=\ket{\chi_i}\bra{\eta_i}$, where
$\ket{\eta_i}$ is an eigenvector of
$({X_0}^\dagger)^{-1}{X_1}^\dagger$. If
$({X_0}^\dagger)^{-1}{X_1}^\dagger$ has only one eigenvector, then
it is impossible to obtain $\sum_iM_i^\dagger M_i={\mathbb I}$.
Therefore, $({X_0}^\dagger)^{-1}{X_1}^\dagger$ has two eigenvectors.
However, if we can not take them orthogonal to each other, it is
again impossible to have $\sum_iM_i^\dagger M_i={\mathbb I}$.
Therefore, $({X_0}^\dagger)^{-1}{X_1}^\dagger$ has two orthogonal
eigenvectors, which means that it is diagonalized by the orthonormal
basis, and $M_i$ has to be given as in Eq.(\ref{M2}). The proof for
the case that the Schmidt rank of $\ketab{\psi_1}$ is $2$ is
identical. $\square$

Now we present a Lemma for the cases where both of the basis states
$\{ \ketab{\psi_0},\ketab{\psi_1} \}$ have the Schmidt rank $1$.
\begin{lem}
If the Schmidt rank of both of $\ketab{\psi_0}$ and $\ketab{\psi_1}$
are $1$, then deterministic remote destruction is possible if and
only if $\ket{\psi_0}_{AB},\ket{\psi_1}_{AB}$ are of the form
\begin{eqnarray}
\ket{\psi_0}_{AB}= \ket{a_0}_A\ket{b_0}_B, \ket{\psi_1}_{AB}=
\ket{a_0^\perp}_A\ket{b_0'}_B, \label{s11}
\end{eqnarray}
where $\ket{a_0}_A,\ket{a_0^\perp}_A$ are orthogonal. The
measurement operators $\{M_i\}_i$ are of the form
\[
M_i=\ket{\chi_i}\bra{\eta_i}.
\]
Here, the vector  $\ket{\eta_i}$ have to be $\ket{a_0}$ or
$\ket{a_0^\perp}$ if $\ketb{b_0},\ketb{b_0'}$ are not parallel to
each other, while it can be an arbitrary vector if
$\ketb{b_0},\ketb{b_0'}$ are parallel to each other.
\end{lem}

{\it Proof:} If $\{ \ketab{\psi_0},\ketab{\psi_1} \}$ are of the
forms given by Eq.~(\ref{s11}), then we can take
$\{M_0=\ket{a_0}\bra{a_0}, M_1=\ket{a_0^\perp}\bra{a_0^\perp}\}$ for
destruction. Conversely, suppose that by performing a measurement
represented by $\{M_i\}_i$, deterministic remote destruction is
possible. We represent the basis states as
\[
\ketab{\psi_0}=\keta{f}\ketb{\xi},
\ketab{\psi_1}=\keta{f'}\ketb{\xi'}.
\]
Note that either $\la f,f'\ra=0$ or $\la \xi,\xi'\ra=0$ should hold.
By Lemma \ref{odis}, $M_i$ has to be either of the form {\it 1-3} in
the Lemma. If $\la f,f'\ra \neq 0$, we have  $\la \xi,\xi'\ra=0$ and
the situation {\it 3} cannot occur. Therefore, $M$ has to of the
form $M_i=\ket{\chi_i}\bra{f^\perp}$ or
$M_i=\ket{\chi_i}\bra{{f'}^\perp}$. However, as we have $\la f,f'\ra
\neq 0$, we also have $\la f^\perp,{f'}^{\perp}\ra \neq 0$ and
$M_i$s can not satisfy $\sum_i M_i^\dagger M_i={\mathbb I}$.  Hence
we obtain $\la f,f'\ra =0$. Again by Lemma \ref{odis}, if
$\ketb{\xi}$ and $\ketb{\xi'}$ are not parallel, each $M_i$ has to
be of the form $M_i=\ket{\chi_i}\bra{f}$,
$M_i=\ket{\chi_i}\bra{{f'}}$ while if they are parallel, then
$M_i=\ket{\chi_i}\bra{\eta_i}$ for an arbitrary $\ket{\eta_i}$.
$\square$

Now, let us prove Theorem~\ref{distract} for remote destruction.
From Lemma \ref{matrixform1}, if $\ket{\psi_0}_{AB}$ has Schmidt
rank $2$, destruction is possible if and only if
\[
(X_0^{\dagger})^{-1}X_1^{\dagger}=z_0\ket{f_0}\bra{f_0}+z_1\ket{f_1}\bra{f_1}\equiv
Z^\dagger,\] for some orthogonal basis $f_0,f_1$. This is equivalent
to
\begin{eqnarray}
X_1=ZX_0,\nonumber\\
Z^\dagger Z=ZZ^\dagger. \label{xzx}
\end{eqnarray}
Using orthogonality condition $(b_1,b_0')=(b_0,b_1')=0$, one can
easily check that Eq.~(\ref{apple}) implies Eq.~(\ref{xzx}).
Below, we show that Eq.~(\ref{xzx}) implies Eq.~(\ref{apple}).\\
Now, using Schmidt decomposition, we have
$
X_0= \sqrt{\lambda_0}\ket{a_0} \bra{\bar{b}_0}
+\sqrt{\lambda_1}\ket{a_1} \bra{\bar{b}_1}.
$
Representing $Z$ in this basis as
$
Z=a\ket{a_0} \bra{a_0}+b\ket{a_0} \bra{a_1} +c\ket{a_1}
\bra{a_0}+d\ket{a_1} \bra{a_1},
$
Eq.~(\ref{xzx}) can be written as
$
X_1=ZX_0=\sqrt{\lambda_0}\left( a\ket{a_0}+c \ket{a_1}
\right)\bra{\bar{b}_0} +\sqrt{\lambda_1}\left( b\ket{a_0}+d
\ket{a_1} \right)\bra{\bar{b}_1}.
$
Using the conditions
$
Tr X_1^\dagger X_1=Tr X_0^\dagger X_0,\quad Tr X_0^\dagger X_1=0
$
we obtain
\begin{eqnarray}
\lambda_0(\vert a\vert^2+\vert c\vert^2) +\lambda_1(\vert
b\vert^2+\vert d\vert^2) =\lambda_0+\lambda_1, \label{ichi}
\end{eqnarray}
and
\begin{eqnarray}
\lambda_0 a+ \lambda_1 d=0. \label{ni}
\end{eqnarray}
On the other hand, Eq.~(\ref{xzx}) requires
\begin{eqnarray}
\vert b\vert=\vert c\vert,\quad a\bar b+c\bar d=c \bar a+d\bar b.
\end{eqnarray}
Combining these relations, we see that $a,b,c,d$ have to be of the
form
\begin{eqnarray*}
a=\sqrt{\frac{\lambda_1}{\lambda_0}}\cos\gamma e^{i\alpha},\quad
b=\sin\gamma e^{i\beta},\\ c=\sin\gamma e^{2i\alpha-i\beta},\quad
d=-\sqrt{\frac{\lambda_0}{\lambda_1}}\cos\gamma e^{i\alpha}.
\end{eqnarray*}
Substituting these, $X_1$ is given by
\begin{eqnarray*}
X_1&=&\ket{a_0}\bra{\sqrt\lambda_1\cos \gamma e^{-i\alpha}\ket{\bar
b_0}+ \sqrt\lambda_1\sin \gamma e^{-i\beta}\ket{\bar b_1}} \\
&+&\ket{a_1}\bra{\sqrt\lambda_0\sin \gamma
e^{-2i\alpha+i\beta}\ket{\bar b_0}- \sqrt\lambda_0\cos \gamma
e^{-i\alpha}\ket{\bar b_1}}.
\end{eqnarray*}
Note that $\sqrt\lambda_1\cos \gamma e^{-i\alpha}\ket{\bar b_0}+
\sqrt\lambda_1\sin \gamma e^{-i\beta}\ket{\bar b_1}$ and
$\sqrt\lambda_0\sin \gamma e^{-2i\alpha+i\beta}\ket{\bar b_0}-
\sqrt\lambda_0\cos \gamma e^{-i\alpha}\ket{\bar b_1}$ are
orthogonal. Hence this decomposition in the matrix form gives the
Schmidt decomposition of the vector $\ket{\psi_1}_{AB}$, and we
obtain  Eq.~(\ref{apple}).

\section{Summary and discussions}
\label{summary}

In this paper, we present necessary and sufficient conditions for
remote extraction and destruction of single-qubit information
encoded and spread in two-qubit states.  The conditions show that
there are ways to spread qubit information into a two-qubit Hilbert
space in a less ``non-local'' manner, namely, the recovery
(extraction) of qubit information  at one of the qubits and
irreversible erasure (destruction) of qubit information  can be
achieved by measurements on only one of the qubits and classical
communications, even though the encoded two qubit states are
entangled in general.

The main part of the paper is devoted to the proof of necessity for
remote extraction shown in seven steps. By introducing the notation
of extraction measurements (E-measurements), which are local
measurements preserving orthogonality of the basis states for
encoding and equal probability for measuring each basis state, we
evaluate the conditions of the basis states for the existence of
E-measurements. We have derived necessary and sufficient conditions
for basis states for encoding and the explicit form of measurements.

Necessary and sufficient conditions for remote extraction indicate
the possibility of sharing qubit information between two parties in
an asymmetric manner, namely, we can spread qubit information such
that it can be remotely extracted to only one of the parties by
using LOCC but not to the other party.  Since impossibility of LOCC
tasks implies the existence of a kind of non-locality, the
possibility of asymmetric qubit information sharing indicates that
such non-locality can be asymmetric between two parties. The
obtained necessary and sufficient conditions for remote destruction
indicate that such asymmetric non-locality for irreversibly
destroying qubit information can also be introduced, but the
conditions for asymmetric cases are not compatible with those for
remote extraction. Remote destruction by one of the parties is
possible if and only if remote extraction at that party is possible.

If we consider Alice's system in our model as an environment, it is
a similar situation to {\it quantum lost and found} considered by
Gregoretti and Werner in \cite{QLostandFound}.  In their paper, they
have obtained necessary and sufficient conditions for extracting
quantum information spread over the system and environment due to
the system-environment coupling, by measuring the environment and
performing conditional operations on the system. By using quantum
information theoretical analysis, they have proven that extraction
is possible if and only if when a map $\Lambda$ for the system state
$\ket{\phi}_S$ can be represented by the random unitary channel
$\Lambda (\ket{\phi}_S) = \sum_k {p_k} u_k \ket{\phi}_S \bra{\phi}
u_k^\dagger $, where $\{ u_k\}$ is a set of random unitary operators
acting on the system qubit, and $ p_k$ is probability satisfying
$\sum_{k}{p_k} =1$.

We note that our remote extraction can be regarded as a special case
where both the system and environment consist of a single qubit,
however we consider the conditions for general LOCC whereas they
consider only one-way (from the environment to the system) LOCC for
quantum information extraction. In the case of the single qubit
system-environment, a purification of the map $\Lambda$ gives a
corresponding transformation describing the spread of qubit
information to the two parties, the system and environment;
\begin{eqnarray}
\ket{\alpha e_0+\beta e_1}_S &\rightarrow& \sqrt{p_0} \ket{g_0}_E
\otimes u_0 \ket{\alpha e_0+\beta e_1}_S \nonumber \\
&+& \sqrt{p_1} \ket{g_1}_E \otimes u_1 \ket{\alpha e_0+\beta e_1}_S
\label{Gregoretti}
\end{eqnarray}
for arbitrary $\alpha$ and $\beta$, where $\{ \ket{g_0}_E,
\ket{g_1}_E \}$ is an orthonormal basis of the environment. They are
equivalent to our LOCC extraction conditions for basis states given
by Eqs.~(\ref{cond1}) and (\ref{cond2}).  Therefore, the encoding
conditions for remote extraction of qubit information spread into
two-qubit states are same for both one-way LOCC and general LOCC, in
spite of the inapplicability of the Lo-Popescu theorem to this task.

\begin{acknowledgments}
The authors are grateful to D. Markham for discussions. This work
was supported by the Sumitomo Foundation, the Asahi Glass
Foundation, and the Japan Society of the Promotion of Science. Y. O.
was supported by the COE21 program at Graduate School of
Mathematical Sciences, the University of Tokyo. M. M. was partially
supported by the Special Coordination Funds for Promoting Science
and Technology.
\end{acknowledgments}

\end{document}